\documentclass[11pt]{article}
\usepackage[a4paper, width = 150mm, top = 30mm, bottom = 30mm]{geometry}
\usepackage{jheppub}
\usepackage{mathrsfs}
\usepackage{circuitikz}
\usepackage{float}
\usepackage{pgfplots}
\pgfplotsset{compat=1.18}
\usepackage{subcaption}
\usetikzlibrary{decorations.markings}
\usetikzlibrary{arrows}
\usetikzlibrary{hobby}
\usetikzlibrary{math}
\usetikzlibrary{tikzmark}
\usetikzlibrary{arrows.meta}
\usepackage{ifthen}

\newcommand{\A}{\scalebox{.9}{$\scriptscriptstyle A$}}
\newcommand{\B}{\scalebox{.9}{$\scriptscriptstyle B$}}

\newcommand{\sR}{{}_{R}}
\newcommand{\sL}{{}_{L}}

\newcommand{\Psmall}{\scalebox{.9}{$\scriptscriptstyle P$}}
\newcommand{\F}{\scalebox{.9}{$\scriptscriptstyle F$}}
\newcommand{\Pb}{\bar{\Psmall}}
\newcommand{\Fb}{\bar{\F}}

\newcommand{\calI}{\mathcal{I}}

\newcommand{\bbG}{\mathbb{G}}
\newcommand{\bbGA}{\mathbb{G}_{\text{adv}}}
\newcommand{\bbGR}{\mathbb{G}_{\text{ret}}}

\newcommand{\WR}{W_{\text{ret}}}
\newcommand{\WA}{W_{\text{adv}}}
\newcommand{\Gin}{G^{\text{in}}}
\newcommand{\Kin}{K^{\text{in}}}

\newcommand*\diff{\mathop{}\!\mathrm{d}}

\newcommand{\thetaSK}{\Theta^{{}^{\text{SK}}}}

\newcommand{\Cnst}{\kappa}

\newcommand{\red}[1]{\textcolor{black}{#1}}

\hypersetup{
    colorlinks=true,
    linkcolor=blue,
    filecolor=black,
    citecolor=blue,
    urlcolor=blue,
}

\title{\boldmath An Exterior EFT for Hawking Radiation}
\author{Loganayagam R, Godwin Martin}
\affiliation{International Centre for Theoretical Sciences (ICTS-TIFR)\\ 
Tata Institute of Fundamental Research, Shivakote, Hesaraghatta, Bengaluru 560089, India.}
\emailAdd{nayagam@icts.res.in,godwin.martin@icts.res.in}

\begin{document}

\newcommand{\barbdiode}[2]{
	\draw (#1) -- coordinate[midway](m) (#2);
	\draw[-{Straight Barb[scale=1.5]}] (#1)--(m);
	\draw[{|[scale=2]}-] (m) -- (#2);
}

\newcommand{\capordiordi}[3]{
\ctikzset{capacitors/scale=0.35, 
            diodes/scale=0.35}
\ifthenelse{#1=0}{\draw (#2) to[C] (#3)}{};
\ifthenelse{#1=1}{\barbdiode{#2}{#3}}{};
\ifthenelse{#1=-1}{\barbdiode{#3}{#2}}{};
}

\newcommand{\diordi}[3]{
\ifthenelse{#1=1}{\draw (#2) to[D] (#3)}{};
\ifthenelse{#1=-1}{\draw (#3) to[D] (#2)}{};
}

\newcommand{\threeptvertex}[3]{
\begin{circuitikz}
\ctikzset{capacitors/scale=0.35, 
            diodes/scale=0.35}
\coordinate (a) at (0,1.6);
\coordinate (b) at (-1.5,-0.75);
\coordinate (c) at (1.5,-0.75);
\coordinate (o) at (0,0);
\node at (o) {$\bullet$};
\ifthenelse{#1=0}{\draw (a) to[C] (o);}{}
\ifthenelse{#1=1}{\barbdiode{a}{o}}{}
\ifthenelse{#1=-1}{\barbdiode{a}{o}}
\ifthenelse{#2=0}{\draw (b) to[C] (o);}{}
\ifthenelse{#2=1}{\barbdiode{b}{o}}{}
\ifthenelse{#2=-1}{\barbdiode{b}{o}}
\ifthenelse{#3=0}{\draw (c) to[C] (o);}{}
\ifthenelse{#3=1}{\barbdiode{c}{o}}{}
\ifthenelse{#3=-1}{\barbdiode{c}{o}}
\end{circuitikz}
}

\newcommand{\fourptkel}[5]{
\begin{circuitikz}[scale=0.5]
\ctikzset{capacitors/scale=0.5, 
            diodes/scale=0.5}
\coordinate (A) at (-4,0);
\coordinate (B) at (4,0);
\coordinate (J1) at (-3.5,0);
\coordinate (J2) at (-1.5,0);
\coordinate (J3) at (1.5,0);
\coordinate (J4) at (3.5,0);
\coordinate (z1) at (-2.5,-2.5);
\coordinate (z2) at (2.5,-2.5);
\node at ($ (J1) + (0,0.5) $) {\ifthenelse{#1=1}{\small{$J_a(k_1)$}}{\small{$J_d(k_1)$}}};
\node at ($ (J2) + (0,0.5) $) {\ifthenelse{#2=1}{\small{$J_a(k_2)$}}{\small{$J_d(k_2)$}}};
\node at ($ (J3) + (0,0.5) $) {\ifthenelse{#3=1}{\small{$J_a(k_3)$}}{\small{$J_d(k_3)$}}};
\node at ($ (J4) + (0,0.5) $) {\ifthenelse{#4=1}{\small{$J_a(k_4)$}}{\small{$J_d(k_4)$}}};
\node at (z1) {$\bullet$};
\node at (z2) {$\bullet$};
\node at (J1) {$\times$};
\node at (J2) {$\times$};
\node at (J3) {$\times$};
\node at (J4) {$\times$};
\draw[dashed] (A)--(B);
\capordiordi{#1}{J1}{z1}
\capordiordi{#2}{J2}{z1}
\capordiordi{#3}{J3}{z2}
\capordiordi{#4}{J4}{z2}
\capordiordi{#5}{z1}{z2}
\end{circuitikz}
}

\newcommand{\fiveptkel}[7]{
\begin{circuitikz}[scale=0.5]
\ctikzset{capacitors/scale=0.5, 
            diodes/scale=0.5}
\coordinate (A) at (-4,0);
\coordinate (B) at (4,0);
\coordinate (J1) at (-3.5,0);
\coordinate (J2) at (-1.5,0);
\coordinate (J3) at (0,0);
\coordinate (J4) at (1.5,0);
\coordinate (J5) at (3.5,0);
\coordinate (z1) at (-2.5,-2.5);
\coordinate (z2) at (0,-2.5);
\coordinate (z3) at (2.5,-2.5);
\node at ($ (J1) + (0,0.5) $) {\ifthenelse{#1=1}{\tiny{$J_a(k_1)$}}{\tiny{$J_d(k_1)$}}};
\node at ($ (J2) + (0,0.5) $) {\ifthenelse{#2=1}{\tiny{$J_a(k_2)$}}{\tiny{$J_d(k_2)$}}};
\node at ($ (J3) + (0,0.5) $) {\ifthenelse{#3=1}{\tiny{$J_a(k_3)$}}{\tiny{$J_d(k_3)$}}};
\node at ($ (J4) + (0,0.5) $) {\ifthenelse{#4=1}{\tiny{$J_a(k_4)$}}{\tiny{$J_d(k_4)$}}};
\node at ($ (J5) + (0,0.5) $) {\ifthenelse{#5=1}{\tiny{$J_a(k_5)$}}{\tiny{$J_d(k_5)$}}};
\node at (z1) {$\bullet$};
\node at (z2) {$\bullet$};
\node at (z3) {$\bullet$};
\node at (J1) {$\times$};
\node at (J2) {$\times$};
\node at (J3) {$\times$};
\node at (J4) {$\times$};
\node at (J5) {$\times$};
\draw[dashed] (A)--(B);
\capordiordi{#1}{J1}{z1}
\capordiordi{#2}{J2}{z1}
\capordiordi{#3}{J3}{z2}
\capordiordi{#4}{J4}{z3}
\capordiordi{#5}{J5}{z3}
\capordiordi{#6}{z1}{z2}
\capordiordi{#7}{z2}{z3}
\end{circuitikz}
}

\newcommand{\semicap}[4]{
\draw[thick,{|[scale=1.8]}-] (#1 +  #3,#2 + #4)--(#3,#4);
}

\newcommand{\diodearrow}[4]{
\draw[thick,-{Triangle[scale=1.8,open]}] (#1,#2)--(#1 +  #3 , #2 + #4);
}

\newcommand{\diode}[4]{
	\draw[thick,-{Triangle[scale=1.8,open]}] (#1,#2)--(#1/2 +  #3/2 , #2/2 + #4/2);
	\draw[thick,{|[scale=1.8]}-] (#1/2 +  #3/2,#2/2 + #4/2)--(#3,#4);
}

\newcommand{\threeptPF}[3]{
\begin{tikzpicture}
	\coordinate (z) at (0,-0.5);
	\draw[dashed] (-1.4,1)--(1.4,1);
	\ifthenelse{#1 > 0}{\diode{-1.2}{1}{0}{-0.5}}{\diode{0}{-0.5}{-1.2}{1}};
        \node at (-1.2,1) {$\times$};
	\ifthenelse{#2 > 0}{\diode{0}{1}{0}{-0.5}}{\diode{0}{-0.5}{0}{1}};
        \node at (0,1) {$\times$};
	\ifthenelse{#3 > 0}{\diode{1.2}{1}{0}{-0.5}}{\diode{0}{-0.5}{1.2}{1}};
        \node at (1.2,1) {$\times$};
	\node at (z) {$\bullet$};
        \node at (0,-0.8) {$\zeta$};
\end{tikzpicture}
}

\newcommand{\fourptcontPF}[4]{
\begin{tikzpicture}
	\coordinate (z) at (0,-0.5);
	\draw[dashed] (-1.8,1)--(1.8,1);
	\ifthenelse{#1 > 0}{\diode{-1.6}{1}{0}{-0.5}}{\diode{0}{-0.5}{-1.6}{1}};
        \node at (-1.6,1) {$\times$};
	\ifthenelse{#2 > 0}{\diode{-0.4}{1}{0}{-0.5}}{\diode{0}{-0.5}{-0.4}{1}};
        \node at (-0.4,1) {$\times$};
        \ifthenelse{#3 > 0}{\diode{0.4}{1}{0}{-0.5}}{\diode{0}{-0.5}{0.4}{1}};
        \node at (0.4,1) {$\times$};
	\ifthenelse{#4 > 0}{\diode{1.6}{1}{0}{-0.5}}{\diode{0}{-0.5}{1.6}{1}};
        \node at (1.6,1) {$\times$};
	\node at (z) {$\bullet$};
        \node at (0,-0.8) {$\zeta$};
\end{tikzpicture}
}

\newcommand{\fourptex}[9]{
\begin{tikzpicture}[scale=1.2]
	\draw[dashed] (-1.7,0)--(1.7,0);
	\ifthenelse{#1 > 0}{\diode{-1.5}{0}{-1}{-1}}{\diode{-1}{-1}{-1.5}{0}}
	\node at (-1.5,0) {$\times$};
	\ifthenelse{#3 > 0}{\diode{-0.5}{0}{-1}{-1}}{\diode{-1}{-1}{-0.5}{0}}
	\node at (-0.5,0) {$\times$};
	\ifthenelse{#5 > 0}{\diode{0.5}{0}{1}{-1}}{\diode{1}{-1}{0.5}{0}}
	\node at (0.5,0) {$\times$};
	\ifthenelse{#7 > 0}{\diode{1.5}{0}{1}{-1}}{\diode{1}{-1}{1.5}{0}}
	\node at (1.5,0) {$\times$};
	\ifthenelse{#9 > 0}{\diode{-1}{-1}{1}{-1}}{\diode{1}{-1}{-1}{-1}}
	\node at (-1.1,-1.2) {$\zeta_1$};
	\node at (1.1,-1.2) {$\zeta_2$};
	\node at (-1,-1) {$\bullet$};
	\node at (1,-1) {$\bullet$};
\end{tikzpicture}
}

\newcommand{\fiveptex}[7]{
\begin{tikzpicture}[scale=1]
	\draw[dashed] (-2.3,0)--(2.3,0);
	\ifthenelse{#1 > 0}{\diode{-2}{0}{-1.5}{-1.5}}{\diode{-1.5}{-1.5}{-2}{0}}
	\node at (-2,0) {$\times$};
	\ifthenelse{#2 > 0}{\diode{-1}{0}{-1.5}{-1.5}}{\diode{-1.5}{-1.5}{-1}{0}}
	\node at (-1,0) {$\times$};
	\ifthenelse{#3 > 0}{\diode{0}{0}{0}{-1.5}}{\diode{0}{-1.5}{0}{0}}
	\node at (0,0) {$\times$};
	\ifthenelse{#4 > 0}{\diode{1}{0}{1.5}{-1.5}}{\diode{1.5}{-1.5}{1}{0}}
	\node at (1,0) {$\times$};
	\ifthenelse{#5 > 0}{\diode{2}{0}{1.5}{-1.5}}{\diode{1.5}{-1.5}{2}{0}}
	\node at (2,0) {$\times$};
	\ifthenelse{#6 > 0}{\diode{-1.5}{-1.5}{0}{-1.5}}{\diode{0}{-1.5}{-1.5}{-1.5}}
	\node at (-1.6,-1.7) {$\zeta_1$};
	\node at (0.1,-1.7) {$\zeta_2$};
	\node at (-1.5,-1.5) {$\bullet$};
	\node at (0,-1.5) {$\bullet$};
	\ifthenelse{#7 > 0}{\diode{0}{-1.5}{1.5}{-1.5}}{\diode{1.5}{-1.5}{0}{-1.5}}
	\node at (1.6,-1.7) {$\zeta_3$};
	\node at (1.5,-1.5) {$\bullet$};
\end{tikzpicture}
}

\newcommand{\sixptlinex}[9]{
\begin{tikzpicture}[scale=1]
	\draw[dashed] (-2.8,0)--(2.8,0);
	\ifthenelse{#1 > 0}{\diode{-2.5}{0}{-2}{-1.5}}{\diode{-2}{-1.5}{-2.5}{0}}
	\node at (-2.5,0) {$\times$};
	\ifthenelse{#2 > 0}{\diode{-1.5}{0}{-2}{-1.5}}{\diode{-2}{-1.5}{-1.5}{0}}
	\node at (-1.5,0) {$\times$};
	\ifthenelse{#3 > 0}{\diode{-0.5}{0}{-0.5}{-1.5}}{\diode{-0.5}{-1.5}{-0.5}{0}}
	\node at (-0.5,0) {$\times$};
	\ifthenelse{#4 > 0}{\diode{0.5}{0}{0.5}{-1.5}}{\diode{0.5}{-1.5}{0.5}{0}}
	\node at (0.5,0) {$\times$};
	\ifthenelse{#5 > 0}{\diode{1.5}{0}{2}{-1.5}}{\diode{2}{-1.5}{1.5}{0}}
	\node at (1.5,0) {$\times$};
	\ifthenelse{#6 > 0}{\diode{2.5}{0}{2}{-1.5}}{\diode{2}{-1.5}{2.5}{0}}
	\node at (2.5,0) {$\times$};
	\ifthenelse{#7 > 0}{\diode{-2}{-1.5}{-0.5}{-1.5}}{\diode{-0.5}{-1.5}{-2}{-1.5}}
	\node at (-2,-1.5) {$\bullet$};
	\node at (-0.5,-1.5) {$\bullet$};
	\ifthenelse{#8 > 0}{\diode{-0.5}{-1.5}{0.5}{-1.5}}{\diode{0.5}{-1.5}{-0.5}{-1.5}}
	\node at (0.5,-1.5) {$\bullet$};
	\ifthenelse{#9 > 0}{\diode{0.5}{-1.5}{2}{-1.5}}{\diode{2}{-1.5}{0.5}{-1.5}}
	\node at (0.5,-1.5) {$\bullet$};
	\node at (2,-1.5) {$\bullet$};
\end{tikzpicture}
}

\newcommand{\sixptmercex}[9]{
\begin{tikzpicture}[scale=1]
	\draw[dashed] (-2.8,0)--(2.8,0);
	\ifthenelse{#1 > 0}{\diode{-2.5}{0}{-2}{-3}}{\diode{-2}{-3}{-2.5}{0}}
	\node at (-2.5,0) {$\times$};
	\ifthenelse{#2 > 0}{\diode{-1.5}{0}{-2}{-3}}{\diode{-2}{-3}{-1.5}{0}}
	\node at (-1.5,0) {$\times$};
	\ifthenelse{#3 > 0}{\diode{-0.5}{0}{0}{-1.25}}{\diode{0}{-1.25}{-0.5}{0}}
	\node at (-0.5,0) {$\times$};
	\ifthenelse{#4 > 0}{\diode{0.5}{0}{0}{-1.25}}{\diode{0}{-1.25}{0.5}{0}}
	\node at (0.5,0) {$\times$};
	\ifthenelse{#5 > 0}{\diode{1.5}{0}{2}{-3}}{\diode{2}{-3}{1.5}{0}}
	\node at (1.5,0) {$\times$};
	\ifthenelse{#6 > 0}{\diode{2.5}{0}{2}{-3}}{\diode{2}{-3}{2.5}{0}}
	\node at (2.5,0) {$\times$};
	\ifthenelse{#7 > 0}{\diode{-2}{-3}{0}{-2.5}}{\diode{0}{-2.5}{-2}{-3}}
	\node at (-2,-3) {$\bullet$};
	\node at (0,-2.5) {$\bullet$};
	\ifthenelse{#8 > 0}{\diode{0}{-1.25}{0}{-2.5}}{\diode{0}{-2.5}{0}{-1.25}}
	\node at (0,-1.25) {$\bullet$};
	\ifthenelse{#9 > 0}{\diode{0}{-2.5}{2}{-3}}{\diode{2}{-3}{0}{-2.5}}
	\node at (2,-3) {$\bullet$};
\end{tikzpicture}
}


\newcommand{\barbdiodeOTO}[3]{
	\draw[#3] #1 -- coordinate[midway](m) #2;
	\draw[#3,-{Straight Barb[scale=1.5]}] #1--(m);
	\draw[#3,{|[scale=2]}-] (m) -- #2;
}
\newcommand{\colordiordiorcap}[3]{
\ifthenelse{#1=111}{\draw #2 to[C] #3}{};
\ifthenelse{#1=101}{\barbdiodeOTO{#2}{#3}{black}}{};
\ifthenelse{#1=110}{\barbdiodeOTO{#3}{#2}{black}}{};
\ifthenelse{#1=211}{\draw[magenta] #2 to[C] #3}{};
\ifthenelse{#1=201}{\barbdiodeOTO{#2}{#3}{magenta}}{};
\ifthenelse{#1=210}{\barbdiodeOTO{#3}{#2}{magenta}}{};
}

\newcommand{\threept}[3]{
\begin{circuitikz}
\ctikzset{capacitors/scale=0.4, 
            diodes/scale=0.4}
\coordinate (J1) at (-1.2,1);
\coordinate (J2) at (0,1);
\coordinate (J3) at (1.2,1);
\coordinate (z) at (0,-0.5);
\draw[dashed] (-1.4,1)--(1.4,1);
\colordiordiorcap{#1}{(J1)}{(z)};
\colordiordiorcap{#2}{(J2)}{(z)};
\colordiordiorcap{#3}{(J3)}{(z)};
\node at (z) {$\bullet$};
\node at ($ (J1) + (0,0.3) $) {\ifthenelse{#1=101}{\small{$J_a(k_1)$}}{\small{$J_d(k_1)$}}};
\node at ($ (J2) + (0,0.3) $) {\ifthenelse{#2=101}{\small{$J_a(k_2)$}}{\small{$J_d(k_2)$}}};
\node at ($ (J3) + (0,0.3) $) {\ifthenelse{#3=101}{\small{$J_a(k_3)$}}{\small{$J_d(k_3)$}}};
\end{circuitikz}
}

\newcommand{\threeptOTO}[8]{
\begin{tikzpicture}[#5]
\coordinate (O) at (0,0);
\coordinate (A) at (0,1.6);
\coordinate (B) at (-1.5,-0.75);
\coordinate (C) at (1.5,-0.75);
\ifthenelse{#6=0}{\draw[{Straight Barb[#1]}-, #2, thick] (A)--(O)}{\draw[{|[#1]}-, #2, thick] (A)--(O)};
\ifthenelse{#7=0}{\draw[{Straight Barb[#1]}-, #3, thick] (B)--(O)}{\draw[{|[#1]}-, #3, thick] (B)--(O)};
\ifthenelse{#8=0}{\draw[{Straight Barb[#1]}-, #4, thick] (C)--(O)}{\draw[{|[#1]}-, #4, thick] (C)--(O)};
\node at (O) {$\bullet$};
\end{tikzpicture}
}

\newcommand{\fourptOTO}[9]{
\begin{tikzpicture}[scale=0.5]
\coordinate (O) at (0,0);
\coordinate (A) at (-1,1);
\coordinate (B) at (-1,-1);
\coordinate (C) at (1,-1);
\coordinate (D) at (1,1);
\ifthenelse{#6=0}{\draw[{Straight Barb[#1]}-, #2, thick] (A)--(O)}{\draw[{|[#1]}-, #2, thick] (A)--(O)};
\ifthenelse{#7=0}{\draw[{Straight Barb[#1]}-, #3, thick] (B)--(O)}{\draw[{|[#1]}-, #3, thick] (B)--(O)};
\ifthenelse{#8=0}{\draw[{Straight Barb[#1]}-, #4, thick] (C)--(O)}{\draw[{|[#1]}-, #4, thick] (C)--(O)};
\ifthenelse{#9=0}{\draw[{Straight Barb[#1]}-, #5, thick] (D)--(O)}{\draw[{|[#1]}-, #4, thick] (D)--(O)};
\node at (O) {$\bullet$};
\end{tikzpicture}
}

\newcommand{\colordiordiorcapkOTO}[3]{
\ifthenelse{#1=-11}{\barbdiodeOTO{#2}{#3}{black}}{};
\ifthenelse{#1=11}{\barbdiodeOTO{#3}{#2}{black}}{};
\ifthenelse{#1=10}{\draw #2 to[C] #3}{};
\ifthenelse{#1=-12}{\barbdiodeOTO{#2}{#3}{magenta}}{};
\ifthenelse{#1=12}{\barbdiodeOTO{#3}{#2}{magenta}}{};
\ifthenelse{#1=-13}{\barbdiodeOTO{#2}{#3}{blue}}{};
\ifthenelse{#1=13}{\barbdiodeOTO{#3}{#2}{blue}}{};
\ifthenelse{#1=-14}{\barbdiodeOTO{#2}{#3}{orange}}{};
\ifthenelse{#1=14}{\barbdiodeOTO{#3}{#2}{orange}}{};
}

\newcommand{\threeptkOTOfull}[3]{
\begin{circuitikz}
\ctikzset{capacitors/scale=0.4, 
            diodes/scale=0.5}
\coordinate (z) at (0,0);
\coordinate (J1) at (-0.9,1.2);
\coordinate (J2) at (0,1.2);
\coordinate (J3) at (0.9,1.2);
\colordiordiorcapkOTO{#1}{(z)}{(J1)};
\colordiordiorcapkOTO{#2}{(z)}{(J2)};
\colordiordiorcapkOTO{#3}{(z)}{(J3)};
\draw[dashed] (-1.1,1.2)--(1.1,1.2);
\node at (z) {\small{$\bullet$}};
\end{circuitikz}
}

\newcommand{\fourpointkOTOcontact}[4]{
\begin{circuitikz}
\ctikzset{capacitors/scale=0.4, 
            diodes/scale=0.5}
\coordinate (z) at (0,-0.5);
\coordinate (J1) at (-1.3,0.8);
\coordinate (J2) at (-0.45,0.8);
\coordinate (J3) at (0.45,0.8);
\coordinate (J4) at (1.3,0.8);
\colordiordiorcapkOTO{#1}{(z)}{(J1)};
\colordiordiorcapkOTO{#2}{(z)}{(J2)};
\colordiordiorcapkOTO{#3}{(z)}{(J3)};
\colordiordiorcapkOTO{#4}{(z)}{(J4)};
\draw[dashed] (-1.5,0.8)--(1.5,0.8);
\node at (z) {\small{$\bullet$}};
\end{circuitikz}
}

\abstract{We propose a general conjecture for evaluating the multiple-discontinuity integrals that appear in real-time holography using gravitational Schwinger-Keldysh (grSK) geometry. Our conjecture is valid for arbitrary non-derivative interactions with any number of bulk tree-level exchanges. It is also consistent with a unitary exterior EFT at finite temperature with correct causal structure. We present a set of Feynman rules underlying this exterior EFT and illustrate it with the computation of four and five-point functions.
}
 \maketitle
\raggedbottom

\section{Introduction}
In the last few years, a new method has emerged which allows to compute  real-time correlators in AdS black holes \cite{Skenderis:2008dg, Skenderis:2008dh, vanRees:2009rw,Glorioso:2018mmw,Chakrabarty:2019aeu,Jana:2020vyx}. This method is based on a contour prescription, dubbed gravitational Schwinger-Keldysh contour(grSK). It automatically gets right various aspects of the boundary real-time correlation functions: unitarity, thermality, locality, and causality.   

The bulk picture presented by the grSK contour is however not entirely clear. The grSK geometry has two copies of the black brane exterior stitched together in a specific way in the near-horizon region. The grSK result for the real-time correlators is naturally given in terms of Witten diagrams in the doubled geometry \cite{Jana:2020vyx}. It is a priori unclear how this doubled geometry picture compares against the physics seen by a bulk observer. A bulk description should be in terms of a QFT in a \emph{single copy} of the spacetime: in fact, we expect a real-time thermal QFT with the associated bulk diagrammatics. The primary message of this work is that, at least for non-derivative interactions, this expectation is indeed met.

For derivative interactions, there is a known complication \cite{Loganayagam:2022teq, Loganayagam:2022zmq} where the grSK prescription gives rise to a specific horizon-localised contribution. For example, the Nambu-Goto action is known to lead to such a horizon contribution\cite{Chakrabarty:2019aeu}. Given that gravity has derivative interactions, understanding the physics behind such contributions is no doubt important. But in this work, we will ignore this subtlety and focus on the physics of the remaining contributions. 

Our goal is to show that, for non-derivative interactions, there is a clear bulk exterior EFT from which real-time correlators can be computed. This is a non-trivial result as it arises from doing multi-contour integrals within Witten diagrams on the grSK geometry and picking up the appropriate multiple-discontinuity (single-discontinuity for contact diagrams, double-discontinuity for single-exchange diagrams, triple-discontinuity for double-exchanges and so on). 

We will give a conjecture for the multiple-discontinuity of an arbitrary tree-level Witten diagram, which we will explicitly check till quadruple-discontinuity (the single-discontinuity was already checked in \cite{Jana:2020vyx} and the double discontinuity was analysed in \cite{Loganayagam:2022zmq}). Assuming this conjecture is true, the laborious procedure of computing multiple-discontinuity integrals over grSK Witten diagrams can be replaced by a simpler single-copy real-time diagrammatics, for non-derivative interactions. For derivative interactions, the full grSK computation seems unavoidable.

This note is organised as follows: we begin section \S\ref{sec:grSK_DerofEFT} by explaining the setup. In section \S\ref{sec:GreenFuncPertSol}, we derive the Green functions and \red{explain the setup of Witten diagrammatics in the grSK spacetime}.  We then present the conjecture for multiple-discontinuity integrals in section \S\ref{sec:MultipleDisc}. In section \S\ref{sec:Diagrammatics}, we present the exterior diagrammatics and compute the influence phase up to five-point functions. We conclude with a summary and future directions in section \S\ref{sec:Discussion}. Much of the technical details regarding the construction of bulk-to-bulk propagators and their properties are \red{relegated} to the appendix.

\section{grSK and derivation of exterior  EFT}\label{sec:grSK_DerofEFT}
We will begin by reviewing the setup of a self-interacting scalar field probing an AdS black-brane. The setup here is similar to the ones described in \cite{Jana:2020vyx,Loganayagam:2022zmq}. Our description here will hence be brief and we will refer the reader to these references for further details.

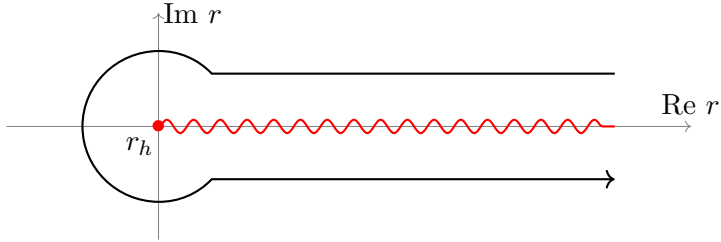
\begin{figure}[H]
\centering
\begin{tikzpicture}
\draw[help lines,->] (-2,0) -- (7,0) coordinate (xaxis);
\draw[help lines,->] (0,-1.5)--(0,1.5) coordinate (yaxis);
\node at (0.45,1.5) {$\text{Im}\ r$};
\node at (7,0.3) {$\text{Re}\ r$};
\draw [thick,->]  (6,0.7) -- (0.7,0.7)
    to [curve through={(0,1)(-1,0) (0,-1)}]
    (0.7,-0.7)--(6,-0.7);
\draw[decoration={snake},decorate, thick, red] (0,0)--(6,0);
\node[red] at (0,0) {$\bullet$};
\node at (-0.25,-0.26) {$r_h$};
\end{tikzpicture}
\caption{\small{The gravitational Schwinger-Keldysh contour on the complex $r$ plane. The red wavy line stands for the branch cut.}}
\label{fig:grSKsaddle}
\end{figure}

Consider a scalar field $\phi$ probing an $\text{AdS}_{d+1}$ Schwarzschild black brane solution. We will adopt the ingoing Eddington-Finkelstein coordinates to write this black brane solution as
\begin{equation}
 \diff s^2 = g_{\A \B}\diff x^{\A} \diff x^{\B}= -r^2 f(r) \diff v^2 + 2 \diff v \diff r + r^2 \diff \mathbf{x}^2 \ ,
\end{equation}
where $f(r) = 1- \frac{r_h^d}{r^d}$ is the emblackening factor with $r_h$ denoting the radius of the horizon. Here $d \mathbf{x}^2$ denotes the line element of a $(d-1)$-dimensional plane. The inverse Hawking temperature $\beta$ of this black brane is given by 
\begin{equation}
\beta = \frac{4\pi}{dr_h} \ .
\end{equation}
Sometimes, we will find it convenient to replace $r$ by a \emph{mock tortoise} coordinate $\zeta$ defined by
\begin{equation}
\frac{\diff r}{\diff \zeta} \equiv \frac{i\beta}{2} r^2 f(r) \ .
\end{equation}
The above metric in $(v,\zeta,\mathbf{x})$ coordinates is given by
\begin{equation}
\diff s^2 = - r^2 f(r) \diff v^2 + i \beta r^2 f(r) \diff v \diff \zeta + r^2 \diff \mathbf{x}^2 \ .
\label{eq:metric(v,zeta,x)}
\end{equation}
The coordinate $\zeta$ is useful in describing the gravitational Schwinger-Keldysh geometry (grSK) we defined earlier \cite{Glorioso:2018mmw, Chakrabarty:2019aeu, Jana:2020vyx}(See Fig.(\ref{fig:grSKsaddle})). Our interest is in studying a scalar field probing this grSK geometry.

For concreteness, we will take the action of the probe scalar $\phi$ to be of the form
\begin{equation}
S = -\int \diff^{d+1}x \ \sqrt{-g} \ \left[\frac{1}{2} \partial_{\A} \phi \partial^{\A} \phi + \frac{\lambda_{3\text{B}}}{3!} \phi^3\right] \ .
\label{eq:ActionPhi3}
\end{equation}
This is a cubic scalar theory on the grSK geometry of the type studied in \cite{Jana:2020vyx,Loganayagam:2022zmq}.  Although our discussion will be centred around scalar $\phi^3$ theory for concreteness, all of the results derived here will hold for a general scalar field theory \red{(with non-derivative interactions)} as discussed in these references.

Before we delve into solving for the field $\phi$ in this spacetime, we note that the above system has the following CFT interpretation: The black brane in the bulk represents the thermal state in the CFT with inverse temperature $\beta$. By the GKPW prescription, the boundary value of $\phi$, which we will denote by $J$ is the source for the dual CFT operator $\mathcal{O}$. The bulk on-shell action evaluated over the grSK geometry is then the generating function\red{al} for Schwinger-Keldysh correlators of $\mathcal{O}$. Alternatively, if $J$ is considered as an external scalar coupled to the thermal CFT bath, the same bulk on-shell action could then be interpreted as \red{the} influence phase of the external scalar.

\subsection{Green functions and the perturbative solution}\label{sec:GreenFuncPertSol}
We will now review the perturbative solution of this system as described in  \cite{Jana:2020vyx}. We will begin with the equation of motion of the probe scalar which takes the form
\begin{equation}
\nabla^2 \phi = \frac{\lambda_{3\text{B}}}{2!} \phi^2 \ ,
\end{equation}
where $\nabla^2$ is \red{the wave operator} defined as
\begin{equation}
\nabla^2 \equiv \nabla_{\A}\nabla^{\A} \equiv \frac{1}{\sqrt{-g}} \partial_{\B} \Big(\sqrt{-g} g^{\A \B} \partial_{\A} \Big) \ .
\label{eq:WaveOperatorCurvedSpacetime}
\end{equation}
We will find it convenient to pass to the Fourier domain labelled by $\red{p\equiv} (p^0, \mathbf{p})$ dual to the variables $(v,\mathbf{x})$. Therefore, we write
\begin{equation}
\phi (v,\zeta,\mathbf{x}) \equiv \int_p \phi ( \zeta, p) e^{-i p^0 v + i \mathbf{p} \cdot \mathbf{x}} \ ,
\label{eq:DefPhiFourierTransform}
\end{equation}
where we have used the notation
\begin{equation}
\int_p \equiv \int \frac{\diff^{d} p}{(2 \pi)^{d}} \ .
\end{equation}
In the Fourier domain, we get a radial ODE for $\phi(\zeta,p)$ of the form
\begin{equation}
\begin{split}
\left(\frac{\diff}{\diff \zeta} + \frac{\beta p^0}{2}\right)& \left[r^{d-1}\left(\frac{\diff}{\diff \zeta} + \frac{\beta p^0}{2}\right) \phi\right] + \left(\frac{i \beta}{2}\right)^2 r^{d-1} \Big((p^0)^2 - \mathbf{p}^2 f(r)\Big) \phi\\
&=\frac{\lambda_{3\text{B}}}{2!} \frac{i\beta}{2} r^{d-1} \frac{\diff r}{\diff \zeta}\   \int_{k_1} \int_{k_2} (2\pi)^d \delta^{(d)}(k_1 + k_2 -p)\ \phi(\zeta, k_1) \phi(\zeta, k_2)\ .
\end{split}
\label{eq:EOMInteractingField}
\end{equation}
Here, by the standard rules of Fourier analysis, the product $\phi^2$ has been converted to a convolution integral in the Fourier domain. This equation can be cast in the notation of \cite{Ghosh:2020lel} using
\begin{equation}
\begin{split}
\mathbb{D}_+ \equiv r^2 f \frac{\diff}{\diff r} - i p^0 &= r^2 f\  \frac{\diff \zeta}{\diff r} \frac{\diff}{\diff \zeta} - i p^0 = \frac{2}{i \beta} \left(\frac{\diff}{\diff \zeta} + \frac{\beta p^0}{2}\right) \ .
\end{split}
\label{eq:DefDPlus}
\end{equation}
We can now solve Eq.\eqref{eq:EOMInteractingField} by perturbatively expanding the solution in increasing orders of $\lambda_{3\text{B}}$ as
\begin{equation}
\phi ( \zeta, {p}) = \phi_{(0)} + \lambda_{3\text{B}} \phi_{(1)} + \lambda^2_{3\rm B} \phi_{(2)} + \ldots \ .
\label{eq:PhiPerturbativeExpansion}
\end{equation}
The equation obeyed by $\phi_{(0)}(\zeta,{p})$ is 
\begin{equation}
\begin{split}
\left(\frac{\diff}{\diff \zeta} + \frac{\beta p^0}{2}\right)& \left[r^{d-1}\left(\frac{\diff}{\diff \zeta} + \frac{\beta p^0}{2}\right) \phi_{(0)}\right] + \left(\frac{i \beta}{2}\right)^2 r^{d-1} \Big((p^0)^2 - \mathbf{p}^2 f(r)\Big) \phi_{(0)} =0 \ .
\end{split}
\label{eq:EOM_Phi0}
\end{equation}

We will denote the ingoing boundary-to-bulk Green function for this system by $G^{\text{in}}(\zeta, {p})$. This Green function is analytic in $\zeta$ (ingoing boundary condition) and satisfies 
\begin{equation}
\lim_{\zeta \to 0} G^{\text{in}}= \lim_{\zeta \to 1} G^{\text{in}} = 1 \ .
\label{eq:LimBoundaryRL_Gin}
\end{equation}
Here, $\zeta=0$ and $\zeta=1$ denote the left- and the right-boundaries of the grSK geometry respectively.
The corresponding outgoing Green function can be obtained via grSK time-reversal to be
\red{
\begin{equation}
    G^{\rm out}(\zeta,p) \equiv e^{-\beta p^0 \zeta} \left[ G^{\rm in}(\zeta,p)\right]^\ast \ .
\end{equation}
}

Given the right-boundary source $J_R(p)$ and the left-boundary source $J_L(p)$, the leading order solution on the grSK contour takes the Son-Teaney form \cite{Son:2009vu,Glorioso:2018mmw,Chakrabarty:2019aeu,Jana:2020vyx}:
\begin{equation}
\phi_{{(0)}}(\zeta,p) = g_{\sR}(\zeta, p)  J_{\sR}(p) - g_{\sL}(\zeta,p) J_{\sL}(p) \ ,
\label{eq:SonTeaneyForm}
\end{equation}
where we have denoted the boundary-to-bulk Green functions from the right and the left boundaries as $g_{\sR}$ and $g_{\sL}$ respectively. Their explicit forms are given by
\begin{equation}
    g_{\sR} (\zeta,p)  \equiv (1+n_p) \left[ G^{\text{in}}(\zeta,p) -  G^{\text{out}} (\zeta,p) \right]
    \label{eq:DefgR}
\end{equation}
and
\begin{equation}
    g_{\sL} (\zeta,p) \equiv n_p  \left[G^{\text{in}}(\zeta,p) - e^{\beta p^0 } G^{\text{out}}(\zeta,p)\right]\ .
    \label{eq:DefgL}
\end{equation}
Here, $n_p$ denotes the Bose-Einstein factor given by
\begin{equation}
n_p \equiv \frac{1}{e^{\beta p^0} -1} \ ,
\end{equation}
where $\beta$ is the inverse Hawking temperature of the black brane. By construction, the Green functions $g_{\sR}$ and $g_{\sL}$ solve Eq.(\ref{eq:EOM_Phi0}) and satisfy the boundary conditions
\begin{equation}
\lim_{\zeta \to 0} g_{\sR} = 0 \ , \qquad \lim_{\zeta \to 0} g_{\sL} = -1 \ ,
\label{eq:LimBoundaryL_gRL}
\end{equation}
as well as
\begin{equation}
\lim_{\zeta \to 1} g_{\sR} = 1 \ , \qquad \lim_{\zeta \to 1} g_{\sL} = 0 \ .
\label{eq:LimBoundaryR_gRL}
\end{equation}

A more convenient form of this solution is written by separating out the ingoing and the outgoing parts, viz., 
\begin{equation}
    \phi_{(0)} (\zeta,p) = -G^{\rm in}(\zeta,p) J_{\Fb}(k) + e^{\beta p^0} G^{\rm out}(\zeta,p) J_{\Pb}(k) \ ,
\end{equation}
where
\begin{equation}
    \begin{split}
        J_{\Fb}(k) &\equiv - \Big[(1+n_{k}) J_{\sR}(k) - n_k J_{\sL}(k)\Big] \ ,\\
        J_{\Pb}(k) &\equiv -n_{k} \Big[J_{\sR}(k) - J_{\sL}(k)\Big] \ .
    \end{split}
    \label{eq:DefBasisPF}
\end{equation}
The $P$ and $F$ here stand for past/future sources. Such a past-future (PF) basis is quite useful\footnote{We alert the reader that we have defined the past-future sources following \cite{Chaudhuri:2018ymp,Jana:2020vyx}. The normalisations we use here differ slightly similar definitions in earlier work, say in \cite{Haehl:2016pec} and \cite{Chou:1984es}.}.

We will now turn to the problem of determining the correction $\phi_{(1)} (\zeta,p)$ to the above leading-order solution. The equation obeyed by $\phi_{(1)} (\zeta,p)$ is 
\begin{equation}
\begin{split}
\left(\frac{\diff}{\diff \zeta} + \frac{\beta p^0}{2}\right)& \left[r^{d-1}\left(\frac{\diff}{\diff \zeta} + \frac{\beta p^0}{2}\right) \phi_{(1)}\right] + \left(\frac{i \beta}{2}\right)^2 r^{d-1} \Big((p^0)^2 - \mathbf{p}^2 f(r)\Big) \phi_{(1)}\\
&\hspace{7cm} = \red{-}\frac{i\beta}{2}  r^{d-1} \frac{\diff r}{\diff \zeta}\mathcal{J}^{\mathfrak{B}}_{(1)}(\zeta,p) \ , \\
\end{split}
\label{eq:EOMPhi1}
\end{equation}
where the bulk source $\mathcal{J}^{\mathfrak{B}}_{(1)}(\zeta,p)$ is given by
\begin{equation}
\mathcal{J}^{\mathfrak{B}}_{(1)}(\zeta,p) \equiv \red{-}\frac{1}{2!}  \int_{k_1} \int_{k_2} (2\pi)^d \delta^{d}(k_1 + k_2 -p)\ \phi_{(0)}(\zeta, k_1) \phi_{(0)}(\zeta, k_2)\ .
\label{eq:Phi1BulkSource}
\end{equation}

We can solve the above equation by constructing a \emph{bulk-to-bulk} Green function denoted by $\mathbb{G} (\zeta| \zeta_0, p)$. Given such a bulk-to-bulk Green function, we can write down the field produced by a bulk source $\mathcal{J}^{\mathfrak{B}}_{(1)}$ as
\begin{equation}
\phi_{(1)} (\zeta,p) = \oint_{\zeta_0} \ \mathbb{G}(\zeta|\zeta_0,p) \ \mathcal{J}^{\mathfrak{B}}_{(1)} (\zeta_0,p) \ .
\label{eq:Phi1FormalExpinBlkBlk}
\end{equation}
Here $\int_{\zeta}$ is defined as
\begin{equation}
\oint_\zeta \equiv \int_{\text{SK}} r^{d-1} \frac{\diff r}{\diff \zeta}\diff \zeta  \ ,
\label{eq:DefIntZeta}
\end{equation}
where $\int_{\text{SK}}$ denotes the integration over the grSK contour extending from $\zeta=0$ (left-boundary at $r=0+i\varepsilon$) to $\zeta=1
$ (right-boundary at $r = 0-i\varepsilon$). This contour encircles the branch cut extending from the horizon radius $r=r_h$ to $r=\infty$. As indicated in the above equation, we denote the position of the bulk source in the grSK contour by $\zeta_0$ and the position where the field is observed by $\zeta$.

The bulk-to-bulk Green function then solves the equation 
\begin{equation}
\begin{split}
&\red{-}\left[\left(\frac{\diff}{\diff \zeta} + \frac{\beta p^0}{2}\right) r^{d-1}\left(\frac{\diff}{\diff \zeta} + \frac{\beta p^0}{2}\right)  + \left(\frac{i \beta}{2}\right)^2 r^{d-1} \Big((p^0)^2 - \mathbf{p}^2 f(r)\Big) \right] \mathbb{G}(\zeta|\zeta_0,p) = \frac{i\beta}{2}\delta(\zeta-\zeta_0) \ .
\label{eq:BlkBlkODE}
\end{split}
\end{equation}
Here $\delta (\zeta- \zeta_0)$ denotes the radial delta-function on the grSK contour. \red{Note that the above equation is just the usual ODE solved by a Green function: $-\nabla^2 \bbG(\zeta|\zeta_0,p) = \frac{\delta(\zeta-\zeta_0)}{\sqrt{-g}}$.}

The above equation should be solved with the following boundary conditions: the bulk-to-bulk Green function $\mathbb{G}(\zeta|\zeta_0,p)$ should be normalisable when $\zeta$ approaches either the right- or the left-boundary, viz.,
\begin{equation}
 \lim_{\zeta \to 0} \mathbb{G}(\zeta|\zeta_0,p) =\lim_{\zeta \to 1} \mathbb{G}(\zeta|\zeta_0,p) = 0 \ .
\label{eq:BlkBlknormcondition}
\end{equation}
This condition implies that $\phi_{(1)}(\zeta,p)$ vanishes at both the boundaries. This in turn ensures that the boundary value of $\phi (\zeta,p)$ continues to be $J_{R}(p)$ on the right-boundary and $J_L(p)$ on the left-boundary respectively.

With the bulk-to-bulk Green function as defined above, we can write down the perturbative solution at any order in the form
\begin{equation}
    \phi_{(n)} (\zeta,p) = \oint_{\zeta_0} \bbG(\zeta|\zeta_0, p) \mathcal{J}^{\mathfrak{B}}_{(n)} (\zeta_0,p) \ ,
\end{equation}
where $\mathcal{J}^{\mathfrak{B}}_{(n)} (\zeta_0,p)$ is the bulk source at the $n$th order in the coupling $\lambda_{3 \rm B}$. These source terms can be expressed purely in terms of the free solution and the bulk-to-bulk Green function. We have already given such an expression for $n=1$ above. For $n=2$, the expression takes the following form
\begin{equation}
    \begin{split}
        \mathcal{J}^{\mathfrak{B}}_{(2)} (\zeta_0,p) &= \red{-}\frac{1}{2!}\int_{k_1}\int_{k_2} (2\pi)^d \delta^{d} (k_1+k_2-p) \left[\phi_{(0)}(\zeta, k_1) \phi_{(1)}(\zeta, k_2) + \phi_{(1)}(\zeta, k_1) \phi_{(0)}(\zeta, k_2)\right]\\
        &= \red{-}\frac{1}{2!} \int_{k_1}\int_{k_2} (2\pi)^d \delta^{d} (k_1+k_2-p)\\
        &\times \oint_{\zeta_0} \Bigg[\phi_{(0)} (\zeta,k_1) \bbG(\zeta|\zeta_0, k_2)\frac{1}{2!}\int_{k_3} \int_{k_4} (2\pi)^d \delta^d(k_3+k_4-k_2) \phi_{(0)}(\zeta_0,k_3) \phi_{(0)}(\zeta_0,k_4)\\
        &\qquad+  \bbG(\zeta|\zeta_0, k_1) \frac{1}{2!} \int_{k_3} \int_{k_4} (2\pi)^d \delta^d(k_3+k_4-k_1) \phi_{(0)}(\zeta_0,k_3) \phi_{(0)}(\zeta_0,k_4) \ \phi_{(0)} (\zeta,k_2) \Bigg]\ .
    \end{split}
\end{equation}
We will now describe how to evaluate the on-shell action over this perturbative solution.

Consider the full action for scalar $\phi^3$ theory in the \red{grSK} geometry, given by
\begin{equation}
    S_{\text{tot}} = S_{\text{bare}} + S_{\text{c.t.}} \ ,
\end{equation}
where $S_{\text{bare}}$ is the bare action defined in Eq.\eqref{eq:ActionPhi3} and $S_{\text{c.t.}}$ constitute the counter-terms. The counter-terms regulate the divergences that appear in the correlation functions of the holographically dual CFT operator $\mathcal{O}$ dual to the bulk field $\phi$. For more details, see \cite{Jana:2020vyx}.

The bare action
\begin{equation}
    S_{\text{bare}} = -\int \diff^{d+1}x \ \sqrt{-g} \ \left[\frac{1}{2} \partial_{\A} \phi \partial^{\A} \phi + \frac{\lambda_{3\text{B}}}{3!} \phi^3\right] \ .
    \end{equation}
\red{can be perturbatively expanded in the bulk coupling strength $\lambda_{3\text{B}}$ as}
\begin{equation}
    S_{\text{bare}} \equiv S = S_{(2)} + S_{(3)} + S_{(4)} + S_{(5)} + \ldots  \ ,
\label{eq:ActionPhi3FormalPertExp}
\end{equation}
where
\begin{equation}
    \begin{split}
        S_{(2)} &\equiv -\frac{1}{2} \int \left(\phi_{(0)} \partial_{\A} \phi_{(0)} \right) \diff \Sigma^{\A}
    \end{split}
\end{equation}
is the leading quadratic contribution, which is a pure boundary term. The correction at any higher order can be obtained by using the expression
\begin{equation}
    S_{\red{\rm bare}}= S_{(2)} -\lambda_{3 \rm B}\int \diff^{d+1} x \ \sqrt{-g}\ \left[-\frac{1}{4} (\phi- \phi_{(0)}) \phi^2 + \frac{\phi^3}{3!}\right] \ ,
\end{equation}
where we have used the equation of motion and the boundary conditions to simplify the answer. This yields the subleading corrections as 
\begin{equation}
    S_{(3)} = -\lambda_{3\text{B}} \int \diff^{d+1} x\ \sqrt{-g}\ \frac{\phi_{(0)}^3}{3!} \ ,
    \label{eq:ActionS3Phi0}
\end{equation}
\begin{equation}
    S_{(4)} = -\frac{\lambda_{3\text{B}}^2}{2} \int \diff^{d+1} x\ \sqrt{-g}\ \frac{\phi_{(0)}^2}{2!} \phi_{(1)}\ ,
    \label{eq:ActionS4Phi0Phi1}
\end{equation}
\begin{equation}
    S_{(5)} = -\frac{\lambda_{3\text{B}}^3}{2} \int \diff^{d+1} x\ \sqrt{-g}\ \frac{\phi_{(0)}^2}{2!} \phi_{(2)}\ .
    \label{eq:ActionS5Phi0Phi2}
\end{equation}
Here the contribution  $S_{(n)}$ corresponds to the $n$-point influence functional of the probe coupling to the dual CFT. One can check that the algebraic evaluation of this on-shell action over the grSK geometry is equivalent to Witten diagrammatics over that geometry. As described in the introduction, our goal is to now show how such diagrams reduce to integrals over the physical exterior geometry.

\subsection{Multiple-discontinuity in grSK: a conjecture}\label{sec:MultipleDisc}
Once the perturbative solution described above is substituted into the on-shell action, the integration over the grSK geometry involves a contour integral over the complexified radius. \red{The contour of integration encircles} a branch cut of the integrand. Such a contour integral then reduces to a real integral over the discontinuity at that branch cut. This procedure was described in detail by the authors of \cite{Loganayagam:2022zmq}, where they used it to evaluate the exchange diagrams with a single bulk-to-bulk Green function (for vertices corresponding to non-derivative interactions). While their procedure of \red{computing discontinuities} can be \red{repeated} for any tree-level diagram, we will proceed here instead by making an educated guess for the final answer. We will present in this section a conjecture for the multiple-discontinuity that matches with the answer computed via the procedure of \cite{Loganayagam:2022zmq}. The conjecture described below applies to an arbitrary tree-level diagram on the grSK geometry.

\begin{figure}[H]
\centering
\begin{tikzpicture}[decoration={markings, 
    mark= at position 3.5cm with {\arrow{stealth}},
    mark= at position 8.0cm with {\arrow{stealth}},
    mark= at position 0.85 with {\arrow{stealth}}}
]
\node at (0.45,1.5) {$\text{Im}\ r$};
\node at (7,0.3) {$\text{Re}\ r$};
\draw[fill=lightgray!50, postaction={decorate}] [thick] (6,0.7) -- (0.7,0.7)
    to [curve through={(0,1)(-1,0) (0,-1)}]
    (0.7,-0.7)--(6,-0.7);
\draw[help lines,->] (-2,0) -- (7,0) coordinate (xaxis);
\draw[help lines,->] (0,-1.5)--(0,1.5) coordinate (yaxis);
\draw[decoration={snake},decorate, thick, red] (0,0)--(6,0);
\node[red] at (0,0) {$\bullet$};
\node[red] at (6,0) {$\bullet$};
\node at (6.25,-0.26) {$r_c$};
\node at (-0.25,-0.26) {$r_h$};
\node at (4,0.4) {$\mathscr{R}$};
\end{tikzpicture}
\caption{The shaded region $\mathscr{R}$ is the region of analyticity of $\mathscr{F}(\zeta_1, \ldots,\zeta_{n_{\text{v}}})$ in the complex $r$ plane. The red wavy line is the branch cut extending from the horizon radius $r_h$ to the \red{cutoff} boundary at $r_c$.}
\label{fig:AnalyticStrip}
\end{figure}
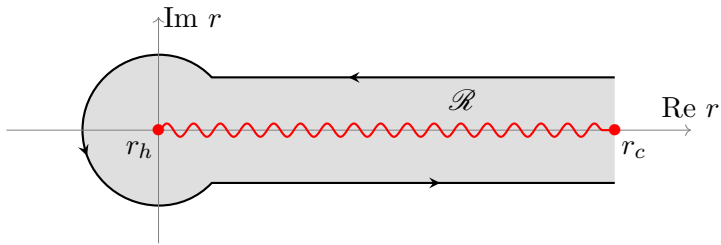

Consider an integral over the \red{grSK} contour in the $r_1,\ r_2, \ldots, \ r_{n_{\text{v}}}$ complex planes. Here $n_{\text{v}}$ stands for the number of bulk vertices that are integrated over. Let us focus only on those diagrams that join these $n_{\text{v}}$ vertices with $n_{\text{e}}$ \red{directed} edges\red{, i.e., $n_{\text{e}}$} internal bulk-to-bulk Green functions. In other words, we are interested in the integral
\begin{equation}
\oint_{\zeta_1} \ldots \oint_{\zeta_{n_\text{v}}} \prod_{i = 1}^{n_\text{v}} e^{\beta \Cnst_i (1- \zeta_i)} \prod_{\ell =1}^{n_{\text{e}}} \bbG(\zeta_{\ell_{f}}|\zeta_{\ell_{i}}, p_{\ell}) \mathscr{F}(\zeta_1, \ldots, \zeta_{n_{\text{v}}}) \ ,
\end{equation}
where $\mathscr{F}(\zeta_1, \ldots, \zeta_{n_{\text{v}}})$ is a function analytic in the strip enclosing the branch cut (see Fig.(\ref{fig:AnalyticStrip})) in the $r_1,\ r_2, \ldots, \ r_{n_{\text{v}}}$ complex planes. \red{Here $\zeta_{\ell_{f}}$ and $\zeta_{\ell_{i}}$ are vertices such that momentum $p_{\ell}$ flows from the latter to the former.}

The above integral over the full grSK contour can be reduced to an integral over just one branch of the contour by evaluating the monodromy around the branch cut. Note that the monodromy arises not only due to the exponential factors that are explicit above, but also due to the theta functions hidden in the bulk-to-bulk propagators. Thus one expects the monodromy factors to be dependent on the temperature $\beta$ as well as the momenta $\Cnst_i$ and $p_\ell$. These factors can be explicitly calculated, but this is a rather tedious calculation. Given this, we present below a conjecture for the values of these factors. 

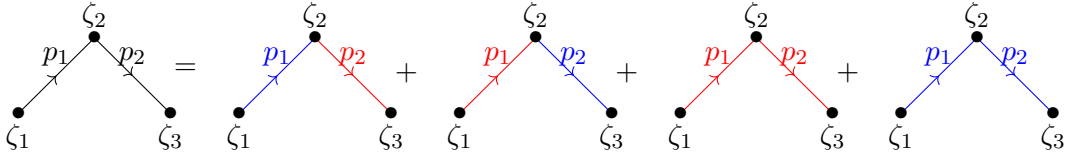
\begin{figure}
\centering
\begin{tikzpicture}
\begin{scope}[decoration={
    markings,
    mark=at position 0.5 with {\arrow{>}}}
    ] 
    \draw[postaction={decorate}] (-1,0)--node[above] {$p_1$}(0,1);
    \draw[postaction={decorate}] (0,1)--node[above] {$p_2$} (1,0);
\end{scope}
\node at (-1,0){$\bullet$};
\node[below] at (-1,0){$\zeta_1$};
\node at (0,1){$\bullet$};
\node[above] at (0,1){$\zeta_2$};
\node at (1,0){$\bullet$};
\node[below] at (1,0){$\zeta_3$};
\node[below] at (1.2,0.8){$=$};
\end{tikzpicture}
\begin{tikzpicture}
\begin{scope}[decoration={
    markings,
    mark=at position 0.5 with {\arrow{>}}}
    ] 
    \draw[blue,postaction={decorate}] (-1,0)--node[above] {$p_1$}(0,1);
    \draw[red,postaction={decorate}] (0,1)--node[above] {$p_2$} (1,0);
\end{scope}
\node at (-1,0){$\bullet$};
\node[below] at (-1,0){$\zeta_1$};
\node at (0,1){$\bullet$};
\node[above] at (0,1){$\zeta_2$};
\node at (1,0){$\bullet$};
\node[below] at (1,0){$\zeta_3$};
\node[below] at (1.2,0.8){$+$};
\end{tikzpicture}
\begin{tikzpicture}
\begin{scope}[decoration={
    markings,
    mark=at position 0.5 with {\arrow{>}}}
    ] 
    \draw[red,postaction={decorate}] (-1,0)--node[above] {$p_1$}(0,1);
    \draw[blue,postaction={decorate}] (0,1)--node[above] {$p_2$} (1,0);
\end{scope}
\node at (-1,0){$\bullet$};
\node[below] at (-1,0){$\zeta_1$};
\node at (0,1){$\bullet$};
\node[above] at (0,1){$\zeta_2$};
\node at (1,0){$\bullet$};
\node[below] at (1,0){$\zeta_3$};
\node[below] at (1.2,0.8){$+$};
\end{tikzpicture}
\begin{tikzpicture}
\begin{scope}[decoration={
    markings,
    mark=at position 0.5 with {\arrow{>}}}
    ] 
    \draw[red,postaction={decorate}] (-1,0)--node[above] {$p_1$}(0,1);
    \draw[red,postaction={decorate}] (0,1)--node[above] {$p_2$} (1,0);
\end{scope}
\node at (-1,0){$\bullet$};
\node[below] at (-1,0){$\zeta_1$};
\node at (0,1){$\bullet$};
\node[above] at (0,1){$\zeta_2$};
\node at (1,0){$\bullet$};
\node[below] at (1,0){$\zeta_3$};
\node[below] at (1.2,0.8){$+$};
\end{tikzpicture}
\begin{tikzpicture}
\begin{scope}[decoration={
    markings,
    mark=at position 0.5 with {\arrow{>}}}
    ] 
    \draw[blue,postaction={decorate}] (-1,0)--node[above] {$p_1$}(0,1);
    \draw[blue,postaction={decorate}] (0,1)--node[above] {$p_2$} (1,0);
\end{scope}
\node at (-1,0){$\bullet$};
\node[below] at (-1,0){$\zeta_1$};
\node at (0,1){$\bullet$};
\node[above] at (0,1){$\zeta_2$};
\node at (1,0){$\bullet$};
\node[below] at (1,0){$\zeta_3$};
\end{tikzpicture}
\caption{An illustration of the multiple-discontinuity formula for the case with two bulk-to-bulk propagators. The leftmost graph above is associated with the combination of bulk-to-bulk propagators in the full contour integral. The directions of the edges as well as the momentum $p_1$ and $p_2$  associated with them are provided by the initial combination of the bulk-to-bulk propagators.  By the conjecture in the text, this graph is replaced by its coloured descendants, the following graphs above.}
\label{fig:GraphsMonodromyIntegrals}
\end{figure}

To present the conjecture, we will require some terminology. Recall that we perform an integral over $n_{\rm v}$ vertices $\zeta_i$. There are $n_{\rm e}$ directed edges connecting these vertices. This is a restatement of the fact that there are $n_{\rm e}$ bulk-to-bulk propagators in the integral. The edges are directed since the bulk-to-propagator has a momentum $p_\ell$ flowing from $\zeta_{{\ell}_{i}}$ to $\zeta_{{\ell}_{f}}$. Thus the integral we wish to evaluate can be represented by a directed graph (see Fig.(\ref{fig:GraphsMonodromyIntegrals})).

According to our conjecture, the full grSK contour integral over the $n_{\rm v}$ vertices reduces to the left-branch integrals over the same vertices. The left-branch integrals have in their integrands certain products of the retarded and the advanced combinations of the bulk-to-bulk Green functions. We will refer the reader to the appendix
\S\ref{sec:Gret&Gadv} for a detailed construction of the left retarded/advanced bulk-to-bulk Green functions: these are bulk-to-bulk Green functions which are normalisable at the left boundary (but not at the right boundary) obeying ingoing/outgoing boundary conditions at the horizon. The explicit forms of these Green functions appear in  Eq.\eqref{eq:BlkBlkretExplicitThetafunc} and Eq.\eqref{eq:BlkBlkadvExplicitThetafunc} respectively. These Green functions were earlier derived in \cite{Arnold:2011hp}.

It is easier to think of the conjecture in terms of the directed graphs we introduced earlier. The original directed graph associated with the full grSK integral splits up into coloured directed graphs as shown in Fig.(\ref{fig:GraphsMonodromyIntegrals}). The blue edges here denote the retarded combination $(-n_{p_\ell})\bbGR(\zeta_{\ell_{f}}|\zeta_{\ell_{i}}, p_{\ell})$ while the red edges denote $(1+n_{p_m})\bbGA(\zeta_{m_{f}}|\zeta_{m_{i}}, p_{m})$, the advanced combination. With these graphs, the conjecture for the multiple-discontinuity can be stated as
\begin{equation}
    \begin{split}
        &\oint_{\zeta_1} \ldots \oint_{\zeta_{n_\text{v}}} \prod_{i = 1}^{n_\text{v}} e^{\beta \Cnst_i (1- \zeta_i)} \prod_{\ell =1}^{n_{\text{e}}} \bbG(\zeta_{\ell_{f}}|\zeta_{\ell_{i}}, p_{\ell}) \mathscr{F}(\zeta_1, \ldots, \zeta_{n_{\text{v}}})\\
        &\qquad= \sum_{\text{graphs}} \int_{\zeta_1} \ldots \int_{\zeta_{n_{\text{v}}}}\prod_{i = 1}^{n_\text{v}} e^{\beta \Cnst_i (1- \zeta_i)} \left[1 - \exp \left(-\beta \Cnst_i +\beta \sum_{j\in \text{out. blue}}p_j-\sum_{j\in \text{in. red}}p_j\right)\right]\\
        &\qquad \quad \times \prod_{\ell = 1}^{n_{\text{e, blue}}} (-n_{p_\ell})\bbG_{\text{ret}}(\zeta_{\ell_{f}}|\zeta_{\ell_{i}}, p_{\ell}) \prod_{m = 1}^{n_{\text{e, red}}} (1+n_{p_m})\bbG_{\text{adv}}(\zeta_{m_{f}}|\zeta_{m_{i}}, p_{m}) \mathscr{F}(\zeta_1, \ldots, \zeta_{n_{\text{v}}}) \ .
    \end{split}
    \label{eq:MultipleDiscontGenFormula}
\end{equation}

The sum outside the integrals on the RHS above is over all possible coloured graphs that can be formed from the original uncoloured graph. The first sum in the exponential runs over edges connected to the $i^{\rm th}$ node that are blue and outgoing (with respect to the $i^{\rm th}$ node). The second sum in the exponential runs over the red edges connected to the $i^{\rm th}$ node that are ingoing (with respect to the $i^{\rm th}$ node). Finally, the products in the last line above are over the blue and the red edges respectively. Here $n_{\text{e, blue}}$ and $n_{\text{e, red}}$ denote the number of blue and red edges respectively. This conjecture for the multiple discontinuity is one of the central results of this work.

In the rest of this subsection, we will illustrate our conjecture with specific examples. We begin  with a general formula for the computation of double integrals of the form
\begin{equation}
\oint_{\zeta_1} \oint_{\zeta_2} e^{\beta \Cnst_1 (1- \zeta_1)} e^{\beta \Cnst_2 (1- \zeta_2)}  \mathscr{F}(\zeta_1,\zeta_2)\ \bbG(\zeta_2|\zeta_1, p)
\end{equation}
over the grSK contour. Such integrals show up in tree diagrams with a single exchange. The general formula is
\begin{equation}
\begin{split}
&\oint_{\zeta_1} \oint_{\zeta_2} e^{\beta \Cnst_1 (1- \zeta_1)} e^{\beta \Cnst_2 (1- \zeta_2)}  \mathscr{F}(\zeta_1,\zeta_2)\ \bbG(\zeta_2|\zeta_1, p) \\
&\hspace{1.5cm}= \int_{\zeta_1} \int_{\zeta_2} e^{\beta \Cnst_1 (1- \zeta_1)} e^{\beta \Cnst_2 (1- \zeta_2)}\mathscr{F}(\zeta_1,\zeta_2)\\
&\hspace{2cm}\times \bigg\{-\frac{n_p}{ (1+n_{\Cnst_1-p}) (1+ n_{\Cnst_2})}\ \bbGR(\zeta_2|\zeta_1, p) + \frac{(1+n_p)}{(1+n_{\Cnst_1})(1+ n_{\Cnst_2+p})}\ \bbGA (\zeta_2|\zeta_1, p)\bigg\} \ ,\\
\end{split}
\label{eq:BlkBlkDoublediscGenRes}
\end{equation}
where we have written the monodromy factors $(1- e^{-\beta \Cnst})$ in terms of the Bose-Einstein factor $n_{\Cnst}$ using\footnote{Note that the Bose-Einstein factor $n_k$ satisfies $1+n_k +n_{-k} =0$.}
\begin{equation}
(1- e^{-\beta \Cnst}) = -\frac{1}{n_{-\Cnst}} = \frac{1}{1+ n_\Cnst} \ .
\label{eq:MonodromyFactorBE}
\end{equation}
In particular, we note that for $\Cnst_1 = 0 = \Cnst_2$, we get
\begin{equation}
\oint_{\zeta_1} \oint_{\zeta_2}  \mathscr{F}(\zeta_1,\zeta_2)\ \bbG(\zeta_2|\zeta_1, p) = 0 \ ,
\label{DoublediscAnalVanish}
\end{equation}
for any function $\mathscr{F} (\zeta_1,\zeta_2)$ analytic in the strip enclosing the branch cut in both the $r_1$ and $r_2$ complex planes.

We will now prove this result for the double integral explicitly. Since none of the functions in the integrand has a simple pole at the horizon radius $r_h$, we can ignore the horizon cap in the grSK contour and evaluate the branch-cut discontinuity as
\begin{equation}
    \begin{split}
        &\oint_{\zeta_1} \oint_{\zeta_2} e^{\beta \Cnst_1 (1- \zeta_1)}e^{\beta \Cnst_2 (1- \zeta_2)} \mathscr{F}(\zeta_1,\zeta_2)\bbG(\zeta_2|\zeta_1,p)\\
        &\hspace{4cm}= \int_{\zeta_1} \int_{\zeta_2} e^{\beta \Cnst_1 (1- \zeta_1)}e^{\beta \Cnst_2 (1- \zeta_2)} \mathscr{F}(\zeta_1,\zeta_2)\bbG_{\text{dd}}(\zeta_2|\zeta_1,p) \ ,
    \end{split}
\end{equation}
where we have defined $\bbG_{\text{dd}}(\zeta_2|\zeta_1,p)$ as
\begin{equation}
    \begin{split}
        \bbG_{\text{dd}}(\zeta_2|\zeta_1,p) &\equiv \bbG(\zeta_2|\zeta_1,p) - e^{-\beta \Cnst_1} \bbG(\zeta_2|\zeta_1+1,p)\\
        &\hspace{2.3cm} - e^{-\beta \Cnst_2} \bbG(\zeta_2+1,\zeta_1,p) + e^{-\beta (\Cnst_1+ \Cnst_2)} \bbG(\zeta_2+1,\zeta_1+1,p) \ .
    \end{split}
    \label{eq:DefBlkBlkdd}
\end{equation}
The exponential factors in the above expression are simply the monodromy factors due to the branch cut extending from the horizon radius $r_h$ to the cutoff radius $r_c$ on the complex $r$ plane.
We can now use the expression for the bulk-to-bulk propagator given in Eq.\eqref{eq:BlkBlkGinGinst} to evaluate the integrand above explicitly. We will use the theta function identities
\begin{equation}
\thetaSK(\zeta+1>\zeta'+1) = \thetaSK(\zeta'>\zeta) \ . 
\end{equation}
and
\begin{equation}
    \thetaSK(\zeta+1>\zeta') =1 \ , \quad  \quad  \thetaSK(\zeta>\zeta'+1) =0  \ , \  \text{when} \ \zeta, \zeta' <\zeta_h \ .
\end{equation}
Here, $\zeta,\zeta'<\zeta_h$ means that both $\zeta$ and $\zeta'$ are on the \red{upper branch of the} contour. Substituting the definition of $\bbG(\zeta_2|\zeta_1,p)$ from Eq.\eqref{eq:BlkBlkGinGinst} in Eq.\eqref{eq:DefBlkBlkdd}, we get
\begin{equation}
    \begin{split}
        \bbG_{\text{dd}}(\zeta_2|\zeta_1,p) &= \frac{1}{ \left\{ K^{\text{in}}(p) - \left[K^{\text{in}}(p)\right]^\ast \right\}}\\
        &\hspace{1cm} \times \bigg\{ c_1  \ e^{\beta p^0 \zeta_1}\ G^{\text{in}} (\zeta_2, p) G^{\text{in}} (\zeta_1,p) + c_2  \ e^{-\beta p^0 \zeta_2} \left[G^{\text{in}} (\zeta_2,p) G^{\text{in}} (\zeta_1,p)\right]^\ast\\
        &\hspace{2cm}+ c_3 \  \Gin(\zeta_2,p) \left[\Gin(\zeta_1,p)\right]^\ast + c_4  \ e^{\beta p^0(\zeta_1-\zeta_2)} \left[ \Gin(\zeta_2,p)\right]^\ast \Gin(\zeta_1,p) \bigg\} \ ,
    \end{split}
    \label{eq:Gddc1c2c3c4}
\end{equation}
where we have defined
\begin{equation}
    \begin{split}
        c_1 &= n_p \left(1-e^{-\beta (\Cnst_1 - p)}\right)\left(1-e^{-\beta \Cnst_2}\right) \ ,\\
        c_2 &= (1+ n_{p})\left(1-e^{-\beta \Cnst_1}\right)\left(1-e^{-\beta( \Cnst_2+p)}\right) \ ,\\
        c_3 &= -c_2 \thetaSK(\zeta_2<\zeta_1) - c_1  \thetaSK(\zeta_2>\zeta_1) \ ,\\
        c_4 &= -c_2 \thetaSK(\zeta_2>\zeta_1) - c_1  \thetaSK(\zeta_2<\zeta_1) \ .
    \end{split}
    \label{eq:GddCoeffSimplified}
\end{equation}

Substituting these coefficients in Eq.\eqref{eq:Gddc1c2c3c4}, we obtain
\begin{equation}
    \bbG_{\text{dd}}(\zeta_2|\zeta_1,p) =-\frac{n_p}{ (1+n_{\Cnst_1-p}) (1+ n_{\Cnst_2})}\ \bbGR(\zeta_2|\zeta_1, p) + \frac{(1+n_p)}{(1+n_{\Cnst_1})(1+ n_{\Cnst_2+p})}\ \bbGA(\zeta_2|\zeta_1, p) \ ,
    \label{eq:DoubleDiscFormula}
\end{equation}
where we have used the definitions of the retarded and the advanced bulk-to-bulk Green functions from Eq.\eqref{eq:BlkBlkretExplicitThetafunc} and Eq.\eqref{eq:BlkBlkadvExplicitThetafunc} respectively.
Here we have also rewritten the monodromy factors in terms of the Bose-Einstein factor $n_{\Cnst}$ using Eq.\eqref{eq:MonodromyFactorBE}. 

Note that the above answer can also be obtained from the multiple-discontinuity formula we have conjectured above in Eq.~\eqref{eq:MultipleDiscontGenFormula}. In terms of the red and the blue edges used there, the above discontinuity formula can be phrased as
\begin{equation}
    \begin{tikzpicture}
        \begin{scope}[decoration={
            markings,
            mark=at position 0.5 with {\arrow{>}}}
            ] 
            \draw[postaction={decorate}] (-1,0)--node[above] {$p_1$}(1,0);
        \end{scope}
        \node at (1.5,0){$=$};
        \begin{scope}[decoration={
            markings,
            mark=at position 0.5 with {\arrow{>}}}
            ] 
            \draw[blue,postaction={decorate}] (2,0)--node[above] {$p_1$}(4,0);
        \end{scope}
        \node at (4.5,0){$+$};
        \begin{scope}[decoration={
            markings,
            mark=at position 0.5 with {\arrow{>}}}
            ] 
            \draw[red,postaction={decorate}] (5,0)--node[above] {$p_1$}(7,0);
        \end{scope}
        \node at (-1,0){$\bullet$};
        \node[below] at (-1,0){$\zeta_1$};
        \node at (1,0){$\bullet$};
        \node[below] at (1,0){$\zeta_2$};
        \node at (2,0){$\bullet$};
        \node[below] at (2,0){$\zeta_1$};
        \node at (4,0){$\bullet$};
        \node[below] at (4,0){$\zeta_2$};
        \node at (5,0){$\bullet$};
        \node[below] at (5,0){$\zeta_1$};
        \node at (7,0){$\bullet$};
        \node[below] at (7,0){$\zeta_2$};
    \end{tikzpicture}
\end{equation}

We will conclude this subsection by quoting the explicit result for the triple discontinuity. This formula will turn out to be useful to us in computing the five-point influence phase later on. The result is
\begin{equation}
    \begin{split}
        &\oint_{\zeta_1} \oint_{\zeta_2}\oint_{\zeta_2} e^{\beta \Cnst_1 (1- \zeta_1)}e^{\beta \Cnst_2 (1- \zeta_2)}e^{\beta \Cnst_3 (1- \zeta_3)} \mathscr{F}(\zeta_1,\zeta_2, \zeta_3)\bbG(\zeta_3|\zeta_2,p_2)\bbG(\zeta_2|\zeta_1,p_1)\\
        &\hspace{3cm}= \int_{\zeta_1} \int_{\zeta_2} \int_{\zeta_3}  e^{\beta \Cnst_1 (1- \zeta_1)}e^{\beta \Cnst_2 (1- \zeta_2)}e^{\beta \Cnst_3 (1- \zeta_3)} \mathscr{F}(\zeta_1,\zeta_2, \zeta_3) \bbG^{(3)}(\zeta_3|\zeta_2|\zeta_1,p_1,p_2) \ ,
    \end{split}
\end{equation}
where the factor $\bbG^{(3)}(\zeta_3|\zeta_2|\zeta_1,p_1,p_2)$ captures the triple discontinuity across the branch cut and by the general multiple discontinuity formula in Eq.\eqref{eq:MultipleDiscontGenFormula} takes the form
\begin{equation}
    \begin{split}
        \bbG^{(3)}(\zeta_3|\zeta_2|\zeta_1,p_1,p_2) &= -\frac{n_{p_1}n_{p_2}}{(1+n_{\Cnst_1-p_1})(1+n_{\Cnst_2-p_2})(1+n_{\Cnst_3})} \bbGR(\zeta_3|\zeta_2,p_2)\bbGR(\zeta_2|\zeta_1,p_1)\\
        &\quad+\frac{(1+n_{p_1})n_{p_2}}{(1+n_{\Cnst_1})(1+n_{\Cnst_2+p_1-p_2})(1+n_{\Cnst_3})} \bbGR(\zeta_3|\zeta_2,p_2)\bbGA(\zeta_2|\zeta_1,p_1)\\
        &\quad+\frac{n_{p_1}(1+n_{p_2})}{(1+n_{\Cnst_1-p_1})(1+n_{\Cnst_2})(1+n_{\Cnst_3+p_2})} \bbGA(\zeta_3|\zeta_2,p_2)\bbGR(\zeta_2|\zeta_1,p_1)\\
        &\quad-\frac{(1+n_{p_1})(1+n_{p_2})}{(1+n_{\Cnst_1})(1+n_{\Cnst_2+p_1})(1+n_{\Cnst_3+p_2})} \bbGA(\zeta_3|\zeta_2,p_2)\bbGA(\zeta_2|\zeta_1,p_1)\ .
    \end{split}
    \label{eq:TripleDiscFormula}
\end{equation}
This expression shows how the final answer from discontinuities takes the form of the sum over coloured bulk to bulk propagators.

\section{Exterior diagrammatics for the influence phase}\label{sec:Diagrammatics}
Using the multiple discontinuity formula, we can now compute the influence phase at any order in the bulk couplings. In this section, we will present the explicit forms of the three-, four-, and five-point influence phases computed using this conjecture. We have also checked the results quoted below via a direct discontinuity computation.

To present the influence phase in the following, we will \red{employ} the useful notation \cite{Jana:2020vyx}
\begin{equation}
    S_{(n)} = \int \prod_{i=1}^{n} \frac{\diff^d k_i}{(2 \pi)^d}\ (2\pi)^d \delta^{(d)} \left(\sum_{i=1}^{n} k_i\right) \left[\sum_{p=0}^{n}\mathcal{I}_{p,n-p}(k_1, \ldots, k_n)\ \prod_{i=1}^{p} {J}_{\Fb} (k_i) \prod_{j= p+1}^{n} {J}_{\Pb} (k_j)\right] \ .
\label{eq:DefInfluencePhaseNotation}
\end{equation}
In the following, we will quote the explicit results for $\mathcal{I}_{p,n-p}(k_1, \ldots, k_n)$ resulting from grSK. The physical interpretation of $\mathcal{I}_{p,n-p}$ is in terms of a scattering process where $p$ ingoing modes scatter into $(n-p)$ outgoing modes. While the results themselves are somewhat involved, there is an especially illuminating way to summarise them which brings out their physical meaning. We find that our results are succinctly summarised by a diagrammatic expansion governed by the following Feynman rules.

\begin{table}[h]
    \centering
    \begin{tabular}{|c|c|}
 \hline     & \\
       \begin{tikzpicture}[scale=1,baseline=(current bounding box.center)]]
        \diodearrow{0}{0}{0}{1}
        \node at (0,0) {$\bullet$};
        \node at (0.4,0.8) {$k_1$};
        \diodearrow{0}{0}{-1}{-0.5}
        \node at (-1,0) {$k_2$};
        \semicap{1}{-0.5}{0}{0}
        \node at (1,0) {$k_3$};
    \end{tikzpicture}  &  $i \lambda_{3\rm B}  \frac{n_{-k_3}}{n_{k_1+k_2}}$  \\   & \\
    \hline   & \\
    \begin{tikzpicture}[scale=1,baseline=(current bounding box.center)]]
        \diodearrow{0}{0}{0}{1}
        \node at (0,0) {$\bullet$};
        \node at (0.4,0.8) {$k_1$};
        \semicap{-1}{-0.5}{0}{0}
        \node at (-1,0) {$k_2$};
        \semicap{1}{-0.5}{0}{0}
        \node at (1,0) {$k_3$};
        \end{tikzpicture} & $ i \lambda_{3\rm B}  \frac{n_{-k_2} n_{-k_3}}{n_{k_1}} $\\   & \\
    \hline   & \\
    \begin{tikzpicture}[scale=0.8,baseline=(current bounding box.center)]]
        \diodearrow{0}{0}{1}{1}
        \node at (0,0) {$\bullet$};
        \node at (0.9,0.4) {$k_1$};
        \diodearrow{0}{0}{-1}{1}
        \node at (-0.3,0.8) {$k_2$};
        \diodearrow{0}{0}{-1}{-1}
        \node at (-0.8,-0.35) {$k_3$};
        \semicap{1}{-1}{0}{0}
        \node at (0.5,-0.8) {$k_4$};
        \end{tikzpicture} & $  i \lambda_{4\rm B}  \frac{n_{-k_4}}{n_{k_1+k_2+k_3}} $\\   & \\
    \hline   & \\
    \begin{tikzpicture}[scale=0.8,baseline=(current bounding box.center)]]
        \diodearrow{0}{0}{1}{1}
        \node at (0,0) {$\bullet$};
        \node at (0.9,0.4) {$k_1$};
        \diodearrow{0}{0}{-1}{1}
        \node at (-0.3,0.8) {$k_2$};
        \semicap{-1}{-1}{0}{0}
        \node at (-0.8,-0.35) {$k_3$};
        \semicap{1}{-1}{0}{0}
        \node at (0.5,-0.8) {$k_4$};
\end{tikzpicture}    & $i \lambda_{4\rm B}  \frac{n_{-k_3} n_{-k_4}}{n_{k_1+k_2}}$\\  & \\
    \hline   & \\
    \begin{tikzpicture}[scale=0.8,baseline=(current bounding box.center)]]
        \diodearrow{0}{0}{1}{1}
        \node at (0,0) {$\bullet$};
        \node at (0.9,0.4) {$k_1$};
        \semicap{-1}{1}{0}{0}
        \node at (-0.3,0.8) {$k_2$};
        \semicap{-1}{-1}{0}{0}
        \node at (-0.8,-0.35) {$k_3$};
        \semicap{1}{-1}{0}{0}
        \node at (0.5,-0.8) {$k_4$};
        \end{tikzpicture} & $ i \lambda_{4\rm B}  \frac{n_{-k_2}n_{-k_3}n_{-k_4}}{n_{k_1}} $\\   & \\
    \hline
    \end{tabular}
    \caption{Feynman rules for the three-point and the four-point vertices.}
    \label{tab:Vertices}
\end{table}

To obtain the influence phase, one must sum over all possible diagrams with the following Feynman rules:
\begin{enumerate}
    \item Multiply every diagram by $\red{-}i$.
   \item The ingoing/outgoing boundary-to-bulk propagators  are denoted  by 
\begin{equation}
    \begin{tikzpicture}[scale=1.4]
        \draw[dashed] (-0.5,0)--(0.5,0);  
        \diode{0}{0}{0}{-1.7};
        \node at (0,-1.7) {$\bullet$};
        \node at (0.2,-1.7) {$\zeta$};
        \node at (0,0) {$\times$};
        \node at (-0.3,-0.5) {$k$};
        \node at (2.65, -0.75) {$  = (1+n_k)^{-1}\Gin(\zeta,k) J_{\Fb}(k) \ ,$};
    \end{tikzpicture}
\hspace{2cm}
    \begin{tikzpicture}[scale =1.4]
        \draw[dashed] (-0.5,0)--(0.5,0);  
        \diode{0}{-1.7}{0}{0};
        \node at (0,-1.7) {$\bullet$};
        \node at (0.2,-1.7) {$\zeta$};
        \node at (0,0) {$\times$};
        \node at (-0.3,-0.5) {$k$};
        \node at (2.5, -0.75) {$ = G^{\rm out}(\zeta,k ) J_{\Pb}(k) \ .$};
    \end{tikzpicture}
\end{equation}
Here $n_k$ denotes the Bose-Einstein factor associated with the momentum $k$. \red{The momentum always flows from the boundary to the bulk in the boundary-to-bulk propagators.}
  \item The retarded bulk-to-bulk propagator (normalisable at $\zeta=0$)  is given by 
\begin{equation}
    \begin{tikzpicture}[scale=1.4]
        \diode{-1}{0}{1}{0};
        \node at (-0.3,-0.2) {$k$};
        \node at (-1,-0.2) {$\zeta_1$};
        \node at (1,-0.2) {$\zeta_2$};
        \node at (-1,0) {$\bullet$};
        \node at (1,0) {$\bullet$};
        \node at (2.5,0) {$ = -i \bbGR (\zeta_2|\zeta_1,k) \ .$};
    \end{tikzpicture}
\end{equation}
\red{In the bulk-to-bulk propagator, the momentum flows from the left to the right.}
\item The Feynman rules for the vertices are given in table~\ref{tab:Vertices}. Here all the momenta are taken to be ingoing \red{at the vertex}. We have also included here the vertex rules in the presence of $\phi^4$ interactions to show the explicit thermal structure of these vertices. For an $n$-point vertex, the temperature dependence is given by
\begin{equation}
    \frac{\prod_{i \in |} n_{-k_i}}{ n_{k_\triangleright}} \ , \quad \text{where} \quad k_{\triangleright} \equiv \sum_{j \in\ \triangleright} k_j\ .
\end{equation}
Here, $|$ denotes the set of all semi-propagators emanating from that vertex and ending in a $|$. Similarly, $\triangleright$ denotes the set of all semi-propagators emanating from that vertex and ending in a $\triangleright$. This is a well-known structure of vertices in real-time finite temperature diagrammatics\cite{vanEijck:1992mq, vanEijck:1994rw, Gelis:1997zv, Carrington:2006xj, Gelis:2019yfm}.
\item The vertices are integrated over the exterior of the black brane. We will denote the radial exterior integral by
        \begin{equation}
            \int_{\rm Ext} \equiv \int_{r_h}^{r_c} \diff r \ r^{d-1} \ .
        \end{equation}
\item As is usual in quantum field theory, the contribution of a given diagram must be weighed by an appropriate symmetry factor.
\end{enumerate}
In what follows, we will present the influence phase expressions along with their associated diagrams.

\subsection{Three/four-point contact influence phase}
\begin{figure}[H]
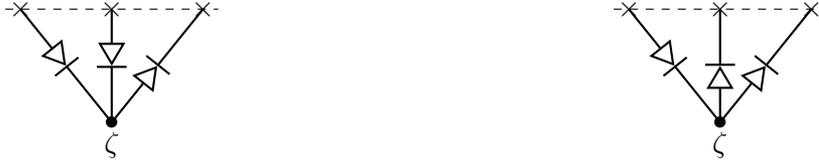

    \centering
    \begin{subfigure}{0.5\textwidth}
        \centering
        \threeptPF{1}{1}{-1}
    \end{subfigure}%
    \begin{subfigure}{0.5\textwidth}
        \centering
        \threeptPF{1}{-1}{-1}
    \end{subfigure}\\
    \caption{Diagrams that contribute to the three-point influence phase.}
    \label{fig:ThreePtIF}
\end{figure}

We will begin with the contact diagrams. The three-point contact diagrams arise by evaluating the on-shell  action contribution of the form \cite{Jana:2020vyx}
\begin{equation}
    S_{(3)} = -\frac{\lambda_{3 \rm B}}{3!} \int \prod_{i=1}^{3} \frac{\diff^d k_i}{(2\pi)^d}\  (2\pi)^d \delta^{(d)} \left(\sum_{i=1}^{3} k_i\right) \oint_{\zeta} \prod_{i=1}^3 \phi_{(0)}(\zeta, k_i) \ .
\end{equation}
The final answer can then be summarised via contact diagrams given in Fig.~(\ref{fig:ThreePtIF}).
The corresponding contribution to the influence phase is 
\begin{equation}\begin{split}
    \mathcal{I}_{2,1} &= \frac{\lambda_{3 \rm B}}{n_{k_3}} \int_{\rm Ext} \frac{1}{2!} \Gin(\zeta,k_1)\Gin(\zeta,k_2)  G^{\rm out}(\zeta,k_3)\ ,\\
    \mathcal{I}_{1,2} &= -\frac{\lambda_{3 \rm B}}{n_{k_2+k_3}} \int_{\rm Ext}  \Gin(\zeta,k_1) \frac{1}{2!} G^{\rm out}(\zeta,k_2)  G^{\rm out}(\zeta,k_3)\ .
\end{split}\end{equation}
Here, the $2!$ factors in the denominator are the symmetry factors of the diagrams. Similarly, the four-point contact diagrams come from \cite{Jana:2020vyx} 
\begin{equation}
    S^{\rm c}_{(4)} = -\frac{\lambda_{4 \rm B}}{4!} \int \prod_{i=1}^{4} \frac{\diff^d k_i}{(2\pi)^d}\  (2\pi)^d \delta^{(d)} \left(\sum_{i=1}^{4} k_i\right) \oint_{\zeta} \prod_{i=1}^4 \phi_{(0)}(\zeta, k_i) \ ,
\end{equation} 
and are given by diagrams of the form
\begin{figure}[H]
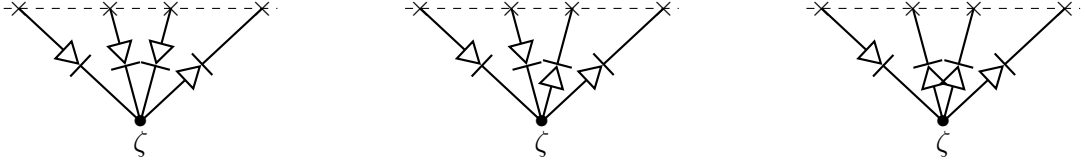

    \centering
    \begin{subfigure}{0.33\textwidth}
        \centering
        \fourptcontPF{1}{1}{1}{-1}
    \end{subfigure}%
    \begin{subfigure}{0.33\textwidth}
        \centering
        \fourptcontPF{1}{1}{-1}{-1}
    \end{subfigure}%
    \begin{subfigure}{0.33\textwidth}
        \centering
        \fourptcontPF{1}{-1}{-1}{-1}
    \end{subfigure}%
    \caption{Contact diagrams that contribute to the four-point influence phase.}
    \label{fig:FourPtContIF}
\end{figure}
The algebraic expressions associated with these diagrams are
\begin{equation}\begin{split}
    \mathcal{I}^{\rm c}_{3,1} &= -\frac{\lambda_{4 \rm B}}{n_{k_4}}\int_{\rm Ext} \frac{1}{3!} \Gin(\zeta,k_1)\Gin(\zeta,k_2)\Gin(\zeta,k_3)  G^{\rm out}(\zeta,k_4)\ ,\\
    \mathcal{I}^{\rm c}_{2,2} &= \frac{\lambda_{4 \rm B}}{n_{k_3+k_4}}\int_{\rm Ext} \frac{1}{2!} \Gin(\zeta,k_1)\Gin(\zeta,k_2) \frac{1}{2!}  G^{\rm out}(\zeta,k_3)  G^{\rm out}(\zeta,k_4)\ ,\\
   \mathcal{I}^{\rm c}_{1,3} &= -\frac{\lambda_{4 \rm B}}{n_{k_2+k_3+k_4}}\int_{\rm Ext}  \Gin(\zeta,k_1) \frac{1}{3!} G^{\rm out}(\zeta,k_2)  G^{\rm out}(\zeta,k_3)  G^{\rm out}(\zeta,k_4)\ .
\end{split}\end{equation}
The reader can explicitly check that all these expressions can be obtained simply by using the Feynman rules quoted above.

\subsection{Four-point exchange influence phase}
We will now move to the four-point exchange diagrams obtained by using the double discontinuity formula Eq.\eqref{eq:DoubleDiscFormula} to evaluate
\begin{equation}
    \begin{split}
        S^{\rm e}_{(4)} &=   \red{-} \int \prod_{i=1}^{4} \frac{\diff^d k_i}{(2\pi)^d}\  (2\pi)^d \delta^{(d)} \left(\sum_{i=1}^{4} k_i\right)\\
        &\times \left(- \frac{\lambda^2_{3\rm B}}{2!}\right)  \oint_{\zeta_1}\oint_{\zeta_2}  \frac{1}{2!}\phi_{(0)} (\zeta_1, k_1) \phi_{(0)} (\zeta_1, k_2) \bbG(\zeta_2|\zeta_1,k_1+k_2) \frac{1}{2!} \phi_{(0)}(\zeta_2, k_3)  \phi_{(0)}(\zeta_2, k_4) \ .
    \end{split}
    \label{eq:Action4PtFourierDomainBlkBlk}
\end{equation}
The resultant answer has many contributions each of which has a diagrammatic representation.  

We will begin with the diagram describing three ingoing modes scattering into one outgoing mode:
\begin{figure}[H]
    \centering
    \fourptex{1}{1}{1}{2}{1}{3}{-1}{4}{1}
    \caption{The diagram that contributes to the coefficient $\calI_{3,1}$}
    \label{fig:4PtFFFP}
\end{figure}
which evaluates to
\begin{equation}
    \begin{split}
        \calI^{\rm e}_{3,1} &= \red{-}\frac{\lambda_{3 \rm B}^2}{n_{k_4}} \int_{\rm Ext_{1,2}}  \frac{1}{2!}\Gin(\zeta_1,k_1) \Gin(\zeta_1,k_2) \ \bbGR(\zeta_2|\zeta_1, k_1+k_2) \ \Gin(\zeta_2,k_3) G^{\rm out}(\zeta_2,k_4) \ .
    \end{split}
\end{equation}
Next is the diagram describing two ingoing modes scattering into two outgoing modes: 
\begin{figure}[H]
    \begin{subfigure}{0.5\textwidth}
        \centering
        \fourptex{1}{1}{1}{2}{-1}{3}{-1}{4}{1}
        \subcaption{}
        \label{fig:4PtFFPP}
    \end{subfigure}%
    \begin{subfigure}{0.5\textwidth}
        \centering
        \fourptex{1}{1}{-1}{3}{1}{2}{-1}{4}{1}
        \subcaption{}
        \label{fig:4PtFPFP1}
    \end{subfigure}
    \caption{The diagrams that contribute to the coefficient $\mathcal{I}_{2,2}$.}
    \label{fig:4PtP2F2}
\end{figure}
The corresponding influence phase expression is 
\begin{equation}
    \begin{split}
        \mathcal{I}^{\rm e}_{2,2} &= \lambda_{3\rm B}^2 \int_{\rm Ext_{1,2}} \\
        &\hspace{1.5cm}\Bigg\{ \frac{1}{n_{k_3+k_4}}    \frac{1}{2!} \Gin(\zeta_1,k_1) \Gin(\zeta_1,k_2) \ \bbGR(\zeta_2|\zeta_1, k_1+k_2)\frac{1}{2!}G^{\rm out}(\zeta_2,k_3) G^{\rm out}(\zeta_2,k_4)\\
        &\hspace{1.3cm}+\frac{1+n_{k_1+k_3}}{(1+n_{k_1})n_{k_4}} \Gin(\zeta_1,k_1)G^{\rm out}(\zeta_1,k_3) \ \bbGR(\zeta_2|\zeta_1, k_1+k_3)\  \Gin(\zeta_2,k_2)  G^{\rm out}(\zeta_2,k_4)\Bigg\} \ .
    \end{split}
\end{equation}

Finally, we come to the last remaining non-zero coefficient $\calI_{1,3}$ describing the scattering of an ingoing mode into three outgoing modes. The only diagram that contributes is given in Fig.(\ref{fig:4PtFPPP}). 
This diagram evaluates to
\begin{equation}
    \begin{split}
        \calI^{\rm e}_{1,3}  &= \red{-}\frac{\lambda_{3\rm B}^2}{n_{k_2+k_3+k_4}}\int_{\rm Ext_{1,2}}  \Gin(\zeta_1,k_1)  G^{\rm out}(\zeta_1,k_2) \ \bbGR(\zeta_2|\zeta_1, k_1+k_2)   \frac{1}{2!} G^{\rm out}(\zeta_2,k_3)  G^{\rm out}(\zeta_2,k_4)  \ .
    \end{split}
    \label{FourPtDiagraI13}
\end{equation}

\begin{figure}[H]
    \centering
    \fourptex{1}{1}{-1}{2}{-1}{3}{-1}{4}{1} 
    \caption{The diagram that contributes to the coefficient $\calI_{1,3}$}
    \label{fig:4PtFPPP}
\end{figure}
Again, all these contributions can be derived immediately from the Feynman rules quoted above.

\subsection{Five-point influence phase}
As a final example, we will evaluate the five-point influence phase which exhibits quite an intricate structure. This is because we are now dealing with a triple discontinuity of the on-shell action, i.e.,  Eq.~\eqref{eq:TripleDiscFormula} applied to the expression
\begin{equation}
    \begin{split}
        S^{\rm e}_{(5)} &= -\frac{\lambda_{3 \rm B}^3}{2!} \int \prod_{i=1}^{5} \frac{\diff^d k_i}{(2\pi)^d}\  (2\pi)^d \delta^{(d)} \left(\sum_{i=1}^{5} k_i\right) \oint_{\zeta_1} \oint_{\zeta_2} \oint_{\zeta_3} \frac{1}{2!}\phi_{(0)} (\zeta_1, k_1) \phi_{(0)} (\zeta_1, k_2)\\
        &\hspace{2cm} \times \bbG(\zeta_2|\zeta_1, k_1+k_2) \ \phi_{(0)} (\zeta_2, k_3) \bbG(\zeta_3|\zeta_2, k_1+k_2+k_3) \frac{1}{2!}\phi_{(0)} (\zeta_3, k_4)\phi_{(0)} (\zeta_3, k_5) \ .
    \end{split}
\end{equation}

We start with the term having one outgoing mode: 
\begin{equation}
    \begin{split}
        \mathcal{I}^{\rm e}_{4,1} &=  \frac{\lambda_{3 \rm B}^3}{n_{k_5}} \int_{\rm Ext_{1,2,3}} \frac{1}{2!} G^{\rm in}(\zeta_1,k_1) G^{\rm in}(\zeta_1,k_2) \bbGR(\zeta_2|\zeta_1,k_1+k_2)\\
        &\hspace{1.5cm}\times  \bigg[ G^{\rm in}(\zeta_2,k_3) \bbGR(\zeta_3|\zeta_2,k_1+k_2+k_3)G^{\rm in}(\zeta_3,k_4) G^{\rm out}(\zeta_3,k_5)\\
        &\hspace{2cm}+ \frac{1}{2!} G^{\rm out}(\zeta_2, k_5) \bbGR(\zeta_2|\zeta_3,k_3+k_4)\frac{1}{2!} G^{\rm in}(\zeta_3,k_4) G^{\rm in}(\zeta_3,k_3) \bigg]   \ ,
    \end{split}
\end{equation}
corresponding to the diagrams
\begin{figure}[H]
    \centering
    \begin{subfigure}{0.33\textwidth}
        \centering
            \fiveptex{1}{1}{1}{1}{-1}{1}{1}
        \caption{}
    \end{subfigure}%
    \begin{subfigure}{0.33\textwidth}
        \centering
            \fiveptex{1}{1}{-1}{1}{1}{1}{-1}
        \caption{}
    \end{subfigure}%
    \caption{Contributions to $\mathcal{I}_{4,1}$}
    \label{fig:FivePtPF41}
\end{figure}

Next, the term with two outgoing modes:  this process gets contributions from five different diagrams:
\begin{equation}
    \begin{split}
        \mathcal{I}^{\rm e}_{3,2} &=  -\lambda_{3 \rm B}^3 \int_{\rm Ext_{1,2,3}} \Bigg\{ \frac{1}{2!} G^{\rm in}(\zeta_1,k_1) G^{\rm in}(\zeta_1,k_2) \bbGR(\zeta_2|\zeta_1,k_1+k_2)\\
        &\times \bigg[G^{\rm in}(\zeta_2,k_3) \bbGR(\zeta_3|\zeta_2,k_1+k_2+k_3) \frac{1}{2!}\frac{1}{n_{k_4+k_5}} G^{\rm out}(\zeta_3,k_4)  G^{\rm out}(\zeta_3,k_5)\\
        &\qquad \quad + G^{\rm out}(\zeta_2, k_4) \bigg(\frac{1+n_{k_1+k_2+k_4}}{1+n_{k_1+k_2}}\frac{1}{n_{k_5}}\bbGR(\zeta_3|\zeta_2,k_1+k_2+k_4)\\
        &\hspace{3cm}+ \frac{1+n_{k_3+k_5}}{1+n_{k_3}}\frac{1}{n_{k_4}} \bbGR(\zeta_2|\zeta_3,k_3+k_5) \bigg)G^{\rm in}(\zeta_3,k_3)  G^{\rm out}(\zeta_3,k_5) \bigg]\\
        & + G^{\rm in}(\zeta_1,k_1) G^{\rm out}(\zeta_1,k_4) G^{\rm in}(\zeta_2,k_3) \bbGR(\zeta_3|\zeta_2, k_1+k_4+k_3) G^{\rm in}(\zeta_3,k_2) G^{\rm out}(\zeta_3,k_5)\\
        &\times \bigg[\frac{1+n_{k_1+k_4}}{1+n_{k_1}}\frac{1}{n_{k_5}}\bbGR(\zeta_2|\zeta_1,k_1+k_4) + \frac{1}{2!} \frac{1+n_{k_1+k_4}}{n_{k_4}}\frac{n_{k_1+k_4+k_3}}{n_{k_3} n_{k_5}}\bbGR(\zeta_1|\zeta_2,k_2+k_3+k_5) \bigg] \Bigg\}\ .
    \end{split}
\end{equation}
\begin{figure}[H]
    \centering
    \begin{subfigure}{0.5\textwidth}
        \centering
            \fiveptex{1}{1}{1}{-1}{-1}{1}{1}
        \caption{}
    \end{subfigure}\\
    \begin{subfigure}{0.5\textwidth}
        \centering
            \fiveptex{1}{1}{-1}{1}{-1}{1}{1}
        \caption{}
    \end{subfigure}%
    \begin{subfigure}{0.5\textwidth}
        \centering
            \fiveptex{1}{1}{-1}{1}{-1}{1}{-1}
        \caption{}
    \end{subfigure}\\
    \begin{subfigure}{0.5\textwidth}
        \centering
            \fiveptex{1}{-1}{1}{1}{-1}{1}{1}
        \caption{}
    \end{subfigure}%
    \begin{subfigure}{0.5\textwidth}
        \centering
            \fiveptex{1}{-1}{1}{1}{-1}{-1}{1}
        \caption{}
    \end{subfigure}\\
    \caption{Contributions to $\mathcal{I}_{3,2}$}
    \label{fig:FivePtPF32}
\end{figure}

\begin{figure}[H]
    \centering
    \begin{subfigure}{0.5\textwidth}
        \centering
            \fiveptex{1}{1}{-1}{-1}{-1}{1}{1}
        \caption{}
    \end{subfigure}\\
    \begin{subfigure}{0.5\textwidth}
        \centering
            \fiveptex{1}{-1}{1}{-1}{-1}{1}{1}
        \caption{}
    \end{subfigure}%
    \begin{subfigure}{0.5\textwidth}
        \centering
            \fiveptex{1}{-1}{1}{-1}{-1}{-1}{1}
        \caption{}
    \end{subfigure}\\
    \begin{subfigure}{0.5\textwidth}
        \centering
            \fiveptex{1}{-1}{-1}{1}{-1}{1}{1}
        \caption{}
    \end{subfigure}%
    \begin{subfigure}{0.5\textwidth}
        \centering
            \fiveptex{1}{-1}{-1}{1}{-1}{1}{-1}
        \caption{}
    \end{subfigure}\\
    \caption{Contributions to $\mathcal{I}_{2,3}$}
    \label{fig:FivePtPF23}
\end{figure}

Next, the term with three outgoing modes, or equivalently the term with two ingoing modes. Just like the previous term, here too there are contributions from five diagrams.
\begin{equation}
    \begin{split}
        \mathcal{I}^{\rm e}_{2,3} &= \lambda_{3 \rm B}^3 \int_{\rm Ext_{1,2,3}}  \Bigg\{\Bigg[ \frac{1}{2!} G^{\rm in}(\zeta_1,k_1) G^{\rm in}(\zeta_1,k_2) \frac{1}{n_{k_3+k_4+k_5}} \bbGR(\zeta_2|\zeta_1,k_1+k_2) G^{\rm out}(\zeta_2, k_3)\\
        &\hspace{2.5cm} + G^{\rm in}(\zeta_1,k_1) G^{\rm out}(\zeta_1,k_3) \bigg(\frac{1+n_{k_1+k_3}}{1+n_{k_1}}\frac{1}{n_{k_4+k_5}}\bbGR(\zeta_3|\zeta_2,k_1+k_2+k_3)\\
        &\hspace{4.5cm}+ \frac{1+n_{k_2+k_4+k_5}}{1+n_{k_2}}\frac{1}{n_{k_3}} \bbGR(\zeta_2|\zeta_3,k_4+k_5) \bigg) G^{\rm in}(\zeta_2,k_2)  \Bigg]\\
        &\hspace{3.5cm} \times  \frac{1}{2!} G^{\rm out}(\zeta_3,k_4)  G^{\rm out}(\zeta_3,k_5)\\
        &\hspace{2.5cm}+ G^{\rm in}(\zeta_1,k_1) G^{\rm out}(\zeta_1,k_4) \bbGR(\zeta_2|\zeta_1,k_1+k_4) G^{\rm out}(\zeta_2,k_3)  G^{\rm in}(\zeta_3,k_2) G^{\rm out}(\zeta_3,k_5)\\
        &\hspace{3.5cm} \times \bigg[\frac{1+n_{k_1+k_3+k_4}}{1+n_{k_1}} \frac{1}{n_{k_5}}\bbGR(\zeta_3|\zeta_2,k_1+k_4+k_3)\\
        &\hspace{4cm}+\frac{1}{2!}\frac{1+n_{k_1+k_4}}{1+n_{k_1}}\frac{1+n_{k_2+k_5}}{1+n_{k_2}}\frac{1}{n_{k_3}}\bbGR(\zeta_2|\zeta_3,k_2+k_5)\bigg] \Bigg\} \ .
    \end{split}
\end{equation}

And finally, we come to the term with four outgoing modes. This term has contributions from two diagrams. 
\begin{equation}
    \begin{split}
        \mathcal{I}^{\rm e}_{1,4} &=  -\frac{\lambda_{3 \rm B}^3}{n_{k_2+k_3+k_4+k_5}} \int_{\rm Ext_{1,2,3}}\bigg[ G^{\rm in}(\zeta_1,k_1) G^{\rm out}(\zeta_1,k_2)  \bbGR(\zeta_2|\zeta_1,k_1+k_2) G^{\rm out}(\zeta_2, k_3)\\
        &\hspace{3.5cm}+ \frac{1}{2!} \frac{1}{2!}G^{\rm out}(\zeta_1,k_3) G^{\rm out}(\zeta_1,k_2)  \bbGR(\zeta_1|\zeta_2,k_1+k_4+k_5) G^{\rm in}(\zeta_2, k_1)\bigg]\\
        &\hspace{5cm}  \times \bbGR(\zeta_3|\zeta_2,k_1+k_2+k_3) \frac{1}{2!} G^{\rm out}(\zeta_3,k_4)  G^{\rm out}(\zeta_3,k_5) \ .
    \end{split}
\end{equation}
\begin{figure}[H]
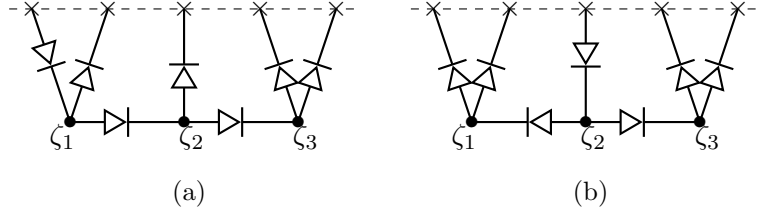

    \centering
    \begin{subfigure}{0.33\textwidth}
        \centering
            \fiveptex{1}{-1}{-1}{-1}{-1}{1}{1}
        \caption{}
    \end{subfigure}%
    \begin{subfigure}{0.33\textwidth}
        \centering
            \fiveptex{-1}{-1}{1}{-1}{-1}{-1}{1}
        \caption{}
    \end{subfigure}
    \caption{Contributions to $\mathcal{I}_{1,4}$}
    \label{fig:FivePtPF14}
\end{figure}
This concludes our discussion of the five-point influence phase. We have also checked our Feynman rules against a few of the six-point diagrams.

\section{Discussion}\label{sec:Discussion}

In this work, we have focused on the real-time dynamics of interacting scalar fields probing the AdS black brane. We have explicitly found the on-shell action of this system evaluated over the gravitational Schwinger-Keldysh (grSK) geometry. This computation is made simple by a conjecture for the multiple-discontinuity integrals (which we have checked up to \red{five}-point functions), as well as a concise set of diagrammatic rules. The emergence of this diagrammatics leads to a simple physical picture behind our answers: the answers given by grSK can simply be reinterpreted in terms of a local unitary EFT living in the exterior of the black brane. The rules governing this EFT are shown to be the familiar ones for a thermal field theory \cite{Gelis:1997zv, Carrington:2006xj, Gelis:2019yfm}.

There are many natural future directions to this work. In particular, it is interesting to study quantum electrodynamics (QED) in the grSK geometry.  QED has a gauge field and an electron field. Gauge fields have already been studied in the grSK geometry in \cite{Ghosh:2020lel,He:2022deg,He:2021jna,Rangamani:2023mok}, where the authors decomposed the gauge field into (designer) scalar fields with a particular non-minimal coupling. Furthermore, interacting spinor fields have been studied in an accompanying work \cite{Martin:2024mdm}. It would now be interesting to put the techniques we have developed, together with these to understand QED in the grSK geometry.

In this work, we have only computed tree-level influence phases in the grSK geometry. Another natural extension, therefore, is the study of loop corrections in the grSK geometry. In particular, one could ask whether the diagrammatic rules we have laid out in the present work extend to the computation of loops. 

One of the outstanding problems in grSK is to ask for a similar geometric contour that would compute the generating functional for out-of-time-ordered correlation (OTOC) functions. Such a contour prescription would be useful in computing the full OTOCs from the AdS black hole bulk and checking that it satisfies all the consistency conditions. It would be interesting to see whether the exterior EFT perspective leads us to such a construction.

\section*{Acknowledgements}
We express our sincere gratitude to Diptarka Das, Alok Laddha, Shiraz Minwalla, Suvrat Raju, Mukund Rangamani, Arnab Rudra, Ashoke Sen, Shivam K Sharma, Omkar Shetye, Aninda Sinha, Akhil Sivakumar, Julio Virrueta, and \red{everyone else in} the ICTS string theory group for valuable discussions. This work was presented at the Strings 2025 Conference at NYU Abu Dhabi and at the Hundred Years of Quantum Mechanics Conference at ICTS-TIFR. GM thanks the organisers of both the conferences for giving him the opportunity to present this work. We acknowledge the support of the Department of Atomic Energy, Government of India, under project no. RTI4001, and the generous support provided by the people of India towards research in the fundamental sciences.

\appendix

\section{Bulk-to-bulk propagators and influence phase}

\subsection{General form of $\mathbb{G}(\zeta|\zeta_0,p)$}\label{sec:BlkBlkGeneralForm}
We will now determine the explicit form of the bulk-to-bulk Green function $\mathbb{G}(\zeta|\zeta_0,p)$. This form has appeared in earlier work \cite{Son:2009vu, Arnold:2011hp} where bulk-to-bulk Green functions are calculated.
When $\zeta \neq \zeta_0$, $\mathbb{G}(\zeta|\zeta_0,p)$ must solve the homogeneous Eq.(\ref{eq:EOM_Phi0}). This implies that $\mathbb{G}(\zeta|\zeta_0,p)$ is a linear combination of the boundary-to-bulk Green functions $g_R(\zeta,p)$ and $g_L(\zeta,p)$ when $\zeta \neq \zeta_0$.

Furthermore, $\mathbb{G}(\zeta|\zeta_0)$ should be normalisable when $\zeta$ approaches either the right or the left boundary, viz.,
\begin{equation}
 \lim_{\zeta \to 0} \mathbb{G}(\zeta|\zeta_0,p) =\lim_{\zeta \to 1} \mathbb{G}(\zeta|\zeta_0,p) = 0 \ .
\end{equation}
This normalisability condition implies that $\mathbb{G}(\zeta|\zeta_0,p)$ is proportional to $g_R(\zeta,p)$ in the region contiguous to the left boundary and is proportional to $g_L (\zeta,p)$ in the region contiguous to the right boundary.

The bulk-to-bulk Green function $\mathbb{G}(\zeta|\zeta_0,p)$ is then fully determined if, in addition to the normalisability condition mentioned above, we also implement an appropriate jump condition at $\zeta = \zeta_0$. The jump condition in question can be derived by integrating Eq.\eqref{eq:BlkBlkODE} in a small interval around $\zeta = \zeta_0$. This yields the condition
\begin{equation}
\left[r^{d-1} \frac{\diff }{\diff \zeta} \mathbb{G}(\zeta|\zeta_0,p)\right]^{\zeta_0^+}_{\zeta_0^-} = \red{-}\frac{i\beta}{2} \ .
\label{eq:BlkBlkJumpCondition}
\end{equation}
Here $\zeta_0^-$ and $\zeta_0^+$ denote the limits as $\zeta$ approaches $\zeta_0$ from the left and the right boundaries respectively.  Thus we conclude that the function $\mathbb{G}(\zeta|\zeta_0,p)$ itself is continuous as we cross $\zeta = \zeta_0$ but its derivative suffers a discontinuity given by the above equality.

We will find it convenient to introduce the following notation for writing down a form for $\mathbb{G}(\zeta|\zeta_0,p)$ that is consistent with the above conditions. To this end, we define
\begin{equation}
\zeta_< \equiv \Bigl\{ 
\begin{array}{cc}
\zeta & \text{if $\zeta$ comes before $\zeta_0$ on the grSK contour}\\
\zeta_0 & \text{if $\zeta_0$ comes before $\zeta$ on the grSK contour} \ ,
\end{array}
\label{eq:DefZetaless}
\end{equation}
and
\begin{equation}
\zeta_> \equiv \Bigl\{ 
\begin{array}{cc}
\zeta & \text{if $\zeta$ comes after $\zeta_0$ on the grSK contour}\\
\zeta_0 & \text{if $\zeta_0$ comes after $\zeta$ on the grSK contour} \ .
\end{array} 
\label{eq:DefZetagreat}
\end{equation}
Using this notation, we can write down a continuous function $\mathbb{G}(\zeta|\zeta_0,p)$ of the form
\begin{equation}
\mathbb{G}(\zeta|\zeta_0,p) = \frac{1}{W(\zeta_0,p)} g_{\sL} (\zeta_>,p) g_{\sR} (\zeta_<,p) \ .
\label{eq:BlkBlkAnsatz}
\end{equation}
The reader can check that this form satisfies the normalisability condition in Eq.\eqref{eq:BlkBlknormcondition}.

\subsubsection{Radial evolution of the Wronskian}\label{sec:WronskianConstancy}
The jump condition in Eq.\eqref{eq:BlkBlkJumpCondition} determines the function $W(\zeta,p)$ to be a Wronskian of the form
\begin{equation}
\begin{split}
W (\zeta,p) &\equiv \red{-}\frac{2}{i\beta}r^{d-1} \left[g_{\sR} \left( \frac{\diff }{\diff \zeta}+\frac{\beta p^0}{2} \right)g_{\sL} - g_{\sL} \left( \frac{\diff }{\diff \zeta}+\frac{\beta p^0}{2} \right) g_{\sR}\right]\\
&= \red{-}g_{\sL}(\zeta,p) \ g_{\sR}^\pi (\zeta,p) \red{+} g_{\sR}(\zeta,p) \ g_{\sL}^\pi(\zeta,p) \ .
\end{split}
\label{eq:WronskianFormalExpression}
\end{equation}
Here $g^\pi_{\sR, \sL} (\zeta,p)$ are the right/left boundary-to-bulk Green functions for the leading-order conjugate field defined via
\begin{equation}
\pi_{(0)}(\zeta,p) \equiv - r^{d-1} \mathbb{D}_+ \phi_{(0)} \equiv g_{\sR}^\pi (\zeta,p) J_{\sR} (p) - g_{\sL}^\pi (\zeta,p) J_{\sL} (p)\ .
\label{eq:ConjugateFieldLeading}
\end{equation}
In AdS/CFT, this conjugate field gets corrected due to counter-term contributions coming from holographic renormalisation as detailed in \cite{Ghosh:2020lel}. Correspondingly, we can define the renormalised version of the above equation as
\begin{equation}
\pi_{(0)}^{\text{ren}}(\zeta,p) \equiv - r^{d-1} \mathbb{D}_+ \phi_{(0)}\Big|_{\text{ren}} \equiv g_{\sR}^{\pi,\ \text{ren}} (\zeta,p) J_{\sR} (p) - g_{\sL}^{\pi,\ \text{ren}}(\zeta,p) J_{\sL} (p)\ .
\label{eq:ConjugateFieldLeadingRen}
\end{equation}

We can equally well replace the boundary-to-bulk Green functions of the conjugate field appearing in the Wronskian with those of the renormalised conjugate field. This is due to the fact that the difference between the bare and the renormalised $g^{\pi}_{\sR, \sL}$ are counter-terms linear in $g_{\sR, \sL}$. Consequently, they cancel out in the Wronskian, thus establishing the claim. Hence, we can write
\begin{equation}
\begin{split}
W(\zeta,p) &= \red{-}g_{\sL}(\zeta,p) \ g_{\sR}^{\pi,\ \text{ren}} (\zeta,p) \red{+} g_{\sR}(\zeta,p) \ g_{\sL}^{\pi,\ \text{ren}}(\zeta,p) \ .
\end{split}
\label{eq:WronskianRenormalised}
\end{equation}

To determine the explicit form of this Wronskian, we proceed as follows. We consider the following derivative of the Wronskian 
\begin{equation}
\begin{split}
&\red{-}\frac{i\beta}{2}\left(\frac{\diff}{\diff \zeta} +\beta p^0 \right)W\\
& =  r^{d-1} \left[\left(\frac{\diff}{\diff \zeta} +\frac{\beta p^0}{2} \right) g_{\sR} \times \left( \frac{\diff }{\diff \zeta}+\frac{\beta p^0}{2} \right)g_{\sL} - \left( \frac{\diff }{\diff \zeta}+\frac{\beta p^0}{2} \right)  g_{\sL} \times  \left( \frac{\diff }{\diff \zeta}+\frac{\beta p^0}{2} \right) g_{\sR}\right]\\
& \quad +  g_{\sR}  \left(\frac{\diff}{\diff \zeta} +\frac{\beta p^0}{2} \right) r^{d-1}  \left(\frac{\diff}{\diff \zeta} +\frac{\beta p^0}{2} \right)g_{\sL} -  g_{\sL}  \left(\frac{\diff}{\diff \zeta} +\frac{\beta p^0}{2} \right) r^{d-1}  \left(\frac{\diff}{\diff \zeta} +\frac{\beta p^0}{2} \right)g_{\sR} \ .
\end{split}
\end{equation}
Here we have split the $\beta p^0$ in the LHS equally between the first and the second lines of the RHS. The first line vanishes trivially whereas the second line vanishes by using Eq.(\ref{eq:EOM_Phi0}) satisfied by $g_{\sR}$ and $g_{\sL}$. Consequently we conclude that $e^{\beta p^0 \zeta} \ W(\zeta,p)$ is a constant in $\zeta$.

\subsubsection{Limits of conjugate boundary-to-bulk propagators}
We can use the constancy of $e^{\beta p^0 \zeta} \ W(\zeta,p)$ to determine $W(\zeta,p)$. We will do this by evaluating the combination $e^{\beta p^0 \zeta} \ W(\zeta,p)$ at either one of the boundaries. To do this, we need, in addition to Eq.\eqref{eq:LimBoundaryL_gRL} and Eq.\eqref{eq:LimBoundaryR_gRL}, the boundary limits of $\ g_{{}_{R,L}}^{\pi, \text{ ren}}$. The result we need from \cite{Ghosh:2020lel} is
\begin{equation}
\lim_{\zeta \to \{0,1\}} r^{d-1} \mathbb{D}_+ G^{\text{in}}(\zeta,p)\Big|_{\text{ren}} = \lim_{r \to \infty} r^{d-1} \mathbb{D}_+ G^{\text{in}}(\zeta,p)\Big|_{\text{ren}} \equiv K^{\text{in}}(p) \ ,
\label{eq:GinConjRen}
\end{equation}
where $\mathbb{D}_+$ is the derivative operator defined in \cite{Ghosh:2020lel} which in our variables is given by Eq.\eqref{eq:DefDPlus}.

Furthermore, the function $K^{\text{in}}(p)$ above is the retarded boundary two-point function of the CFT operator $\mathcal{O}$ dual to the scalar $\phi$. The above equality amounts to the Son-Starinets prescription \cite{Son:2002sd} for the retarded boundary two-point functions. According to this prescription, the retarded two-point functions of the CFT can be computed from the ingoing solutions by taking the $r \to \infty$ limit of the ratio of the renormalised conjugate field $\pi_{(0)}^{\text{ren}}$ to the field $\phi_{(0)}$. This reduces to computing the renormalised conjugate field from the ingoing bulk to boundary Green function and subsequently taking the boundary limit as shown above. We will refer the reader to \cite{Ghosh:2020lel} for a proper derivation of this prescription from the grSK contour as well as for the determination of $K^{\text{in}}(p)$ in the small $p$ expansion.

We will also need the time-reversal of the above equation which gives the advanced boundary two-point function from the outgoing solution. We get
\begin{equation}
\lim_{\zeta \to 0} r^{d-1} \mathbb{D}_+ \left[ e^{-\beta p^0 \zeta}(G^{\text{in}})^\ast \right]\Big|_{\text{ren}} = \left(K^{\text{in}}\right)^\ast \ ,
\end{equation}
as well as
\begin{equation}
\lim_{\zeta \to 1} r^{d-1} \mathbb{D}_+ \left[ e^{-\beta p^0 \zeta}(G^{\text{in}})^\ast \right]\Big|_{\text{ren}} = e^{-\beta p^0} (K^{\text{in}})^\ast \ .
\end{equation}
We note that the advanced Green function acquires an additional Boltzmann factor $e^{-\beta p_0}$ at the right boundary. This additional Boltzmann factor can be understood as arising from the monodromy incurred by outgoing solution while traversing the grSK contour. More directly, the above expressions follow from the identity 
\begin{equation}
 \mathbb{D}_+ \left[G^{\text{in}} (\zeta,-p) e^{-\beta p^0 \zeta}\right] = \left[ \mathbb{D}_+ G^{\text{in}} (\zeta,p) \right]_{p \to -p} e^{-\beta p^0 \zeta} \ .
\end{equation} 
 
We can put these expressions together to compute the boundary limits of $g_{\sR, \sL}^{\pi, \ \text{ren}} (\zeta,p)$ via
\begin{equation}
 g_{\sR}^{\pi,\text{ ren}} = -r^{d-1} \mathbb{D}_+ g_{\sR}\Big|_{\text{ren}} \ \quad \text{and} \quad \  g_{\sL}^{\pi,\text{ ren}} = -r^{d-1} \mathbb{D}_+ g_{\sL}\Big|_{\text{ren}} \ .
\end{equation} 
Using the definitions of $g_{\sR, \sL}$ in Eq.\eqref{eq:DefgL} and Eq.\eqref{eq:DefgR}, we get
\begin{equation}
\begin{split}
\lim_{\zeta \to 0} \ g_{\sR}^{\pi,\text{ ren}} &= - (1+ n_p) \left[ K^{\text{in}}- (K^{\text{in}})^\ast\right] \ ,\\
\lim_{\zeta \to 0} \ g_{\sL}^{\pi,\text{ ren}} &= -  n_p \left[ K^{\text{in}} - e^{\beta p^0} (K^{\text{in}})^\ast\right] \ ,
\end{split}
\label{eq:LimBoundaryL_gRLRen}
\end{equation}
as well as
\begin{equation}
\begin{split}
\lim_{\zeta \to 1} \ g_{\sR}^{\pi,\text{ ren}} &= - (1+ n_p) \left[ K^{\text{in}} - e^{-\beta p^0}(K^{\text{in}})^\ast\right] \ ,\\
\lim_{\zeta \to 1} \ g_{\sL}^{\pi,\text{ ren}} &= -  n_p \left[ K^{\text{in}} -  (K^{\text{in}})^\ast\right] \ .
\end{split}
\label{eq:LimBoundaryR_gRLRen}
\end{equation}
These expressions give the Schwinger-Keldysh two-point functions of the dual CFT operator $\mathcal{O}$. This can be gleaned from the expression for the right and the left Schwinger-Keldysh expectation values of $\mathcal{O}$ defined by
\begin{equation}
\langle \mathcal{O}_{\sR} \rangle \equiv \lim_{\zeta \to 1} \pi^{\text{ren}} \ , \quad \ \langle \mathcal{O}_{\sL} \rangle \equiv \lim_{\zeta \to 0} \pi^{\text{ren}} \ .
\label{eq:ExpecvalfromRenConjField}
\end{equation}
A direct evaluation from Eq.\eqref{eq:ConjugateFieldLeadingRen} yields at the leading-order
\begin{equation}
\begin{split}
\langle \mathcal{O}_{\sR} \rangle &= - (1+ n_p) \left[ K^{\text{in}} - e^{-\beta p^0}(K^{\text{in}})^\ast\right]J_{\sR} + n_p \left[ K^{\text{in}} -  (K^{\text{in}})^\ast\right] J_{\sL} +\ldots\ ,\\
\langle \mathcal{O}_{\sL} \rangle &= - (1+ n_p) \left[ K^{\text{in}}- (K^{\text{in}})^\ast\right]J_{\sR} + n_p \left[ K^{\text{in}} - e^{\beta p^0} (K^{\text{in}})^\ast\right] J_{\sL} +\ldots\ .
\end{split}
\label{eq:CFTExpValLeadingRL}
\end{equation}
Here, $\ldots$ denote the contributions from the higher order corrections nonlinear in boundary sources.
As an aside, we note that this can be simplified to
\begin{equation}
\begin{split}
\langle \mathcal{O}_{\sR} \rangle &= K^{\text{in}} J_{\Fb} -\left(K^{\text{in}}\right)^\ast J_{\Pb}+\ldots\ ,\\
\langle \mathcal{O}_{\sL} \rangle &= K^{\text{in}} J_{\Fb} - e^{\beta p^0} \left(K^{\text{in}}\right)^\ast J_{\Pb}+\ldots\ ,
\end{split}
\label{eq:CFTExpValLeadingPF}
\end{equation}
where we have defined the future and the past sources as in the introduction to this report.

\subsubsection{Evaluating the Wronskian}\label{sec:WronskianEvaluation}
In Sec.(\ref{sec:WronskianConstancy}), we have proved the constancy of $e^{\beta p^0 \zeta}$ times the Wronskian. Since we also know the boundary limits of the bulk-to-boundary Green functions (associated with both the field and its conjugate), we can use Eq.\eqref{eq:WronskianRenormalised} to evaluate the value of the Wronskian at the right/left boundaries. Either one of these values along with the constancy of $e^{\beta p^0 \zeta} W(\zeta,p)$ is sufficient to determine the Wronskian everywhere along the grSK contour. Nevertheless, we will now determine both the boundary values and verify that they indeed yield the same result.

The boundary values of $g_{\sR, \sL}$ are as given in Eq.\eqref{eq:LimBoundaryL_gRL} and Eq.\eqref{eq:LimBoundaryR_gRL}, whereas the boundary values of $g_{\sR, \sL}^{\pi, \ \text{ren}}$ are as given in Eq.\eqref{eq:LimBoundaryL_gRLRen} and Eq.\eqref{eq:LimBoundaryR_gRL}. Using these, we obtain at the left-boundary
\begin{equation}
\begin{split}
\lim_{\zeta \to 0} e^{\beta p^0 \zeta} \ W(\zeta,p) &=  \red{-}\lim_{\zeta \to 0}\Big( g_{\sL}\ g_{\sR}^{\pi,\ \text{ren}} - g_{\sR} \ g_{\sL}^{\pi,\ \text{ren}}\Big) =  \red{+} \lim_{\zeta \to 0} \ g_{\sR}^{\pi,\text{ ren}}\\
&  =\red{-}(1+ n_p) \left\{ K^{\text{in}} (p)- \left[K^{\text{in}}(p)\right]^\ast\right\}\ ,
\end{split}
\end{equation}
while at the right-boundary, we get
\begin{equation}
\begin{split}
\lim_{\zeta \to 1} e^{\beta p^0 \zeta} \ W(\zeta,p) &= \red{-}e^{\beta p^0} \lim_{\zeta \to 1} \Big(g_{\sL} \ g_{\sR}^{\pi,\text{ ren}}-g_{\sR} \ g_{\sL}^{\pi, \text{ ren}} \Big) =  \red{+}e^{\beta p^0} \lim_{\zeta \to 1} \ g_{\sL}^{\pi,\text{ ren}}\\
&  = \red{-} (1+ n_p) \left\{ K^{\text{in}} (p)- \left[K^{\text{in}}(p)\right]^\ast\right\}\ .
\end{split}
\end{equation}
We note that both these computations yield the same answer, as expected from the constancy of $e^{\beta p^0 \zeta} W(\zeta,p)$.

The value of the Wronskian on the grSK contour is then determined to be 
\begin{equation}
W(\zeta,p) =  \red{-}(1+ n_p)e^{-\beta p^0 \zeta} \left\{ K^{\text{in}} (p)- \left[K^{\text{in}}(p)\right]^\ast\right\}  \ .
\end{equation}
Substituting this Wronskian into Eq.\eqref{eq:BlkBlkAnsatz}, we obtain
\begin{equation}
\mathbb{G}(\zeta | \zeta_0,p) = \frac{\red{-}e^{\beta p^0 \zeta_0}}{(1+ n_p) \left\{ K^{\text{in}} (p)- \left[K^{\text{in}}(p)\right]^\ast\right\}}  g_{\sL} (\zeta_>,p) g_{\sR} (\zeta_<,p) \ .
\label{eq:BlkBlkWronskianExplicit}
\end{equation}

\subsubsection{Reciprocity of \texorpdfstring{$\mathbb{G}(\zeta|\zeta_0,p)$}{G()}}\label{sec:BlkBlkReciprocity}
Before proceeding further with the analysis, we would find it convenient to prove some useful relations regarding the bulk-to-bulk Green function derived above. By inspection, we note the following relation obeyed by $\mathbb{G}(\zeta|\zeta_0,p)$:
\begin{equation}
e^{\beta p^0 (\zeta - \zeta_0)} \mathbb{G}(\zeta|\zeta_0,p) = \mathbb{G}(\zeta_0|\zeta,p) \ .
\label{eq:BlkBlksourcefieldswitch}
\end{equation}
Another useful relation regarding the bulk-to-bulk Green function is 
\begin{equation}
\mathbb{G}(\zeta_0|\zeta,p) = \mathbb{G}(\zeta|\zeta_0,-p) \ .
\label{eq:BlkBlkreciprocity}
\end{equation}
We will refer to this relation as the \emph{reciprocity relation}.
This can be proved as follows: we reverse the momentum in Eq.\eqref{eq:BlkBlkWronskianExplicit} to get
\begin{equation}
\mathbb{G}(\zeta|\zeta_0,-p) =  \frac{\red{-}e^{-\beta p^0 \zeta_0}}{n_p \left\{K^{\text{in}}(p) - [K^{\text{in}}(p)]^\ast \right\}} g_{\sL}(\zeta_>,-p) g_{\sR} (\zeta_<,-p) \ .
\label{eq:BlkBlkMomRev}
\end{equation}
Here, we have used the relations
\begin{equation}
1+ n_p + n_{-p} = 0 \quad \text{and} \qquad K^{\text{in}} (-p) = \left[K^{\text{in}} (p)\right]^\ast 
\end{equation}
to simplify the denominator.

Using
\begin{equation}
G^{\text{in}} (-p) = \left[G^{\text{in}}(p)\right]^\ast \ ,
\end{equation}
the behaviour of $g_{{}_{R,L}}$ under momentum reversal can be deduced readily. We get
\begin{equation}
\begin{split}
g_{\sR} (\zeta, -p) &= e^{-\beta p^0 (1- \zeta)} g_{\sR} (\zeta, p) \ , \quad \text{and} \quad g_{\sL} (\zeta, -p)= e^{\beta p^0 \zeta} g_{\sL} (\zeta, p) \ .
\end{split}
\label{eq:MomRevgRL}
\end{equation}
The advertised reciprocity relation follows from Eq.\eqref{eq:BlkBlksourcefieldswitch}, Eq.\eqref{eq:BlkBlkMomRev} and Eq.\eqref{eq:MomRevgRL}.

\subsubsection{Expressing $\bbG(\zeta|\zeta_0,p)$ in terms of $\Gin(\zeta,p)$}

In later sections, we will find it convenient to write the bulk-to-bulk Green function $\bbG(\zeta|\zeta_0,p)$ in terms of the ingoing boundary-to-bulk Green function $\Gin(\zeta,p)$ and its complex conjugate. Starting from the expression for the bulk-to-bulk Green function in Eq.\eqref{eq:BlkBlkWronskianExplicit}, we have
\begin{equation}
\begin{split}
\mathbb{G}(\zeta|\zeta_0,p) &= \red{-}\frac{1}{\left\{K^{\text{in}}(p) - \left[K^{\text{in}}(p)\right]^\ast \right\}} \frac{e^{\beta p^0 \zeta_0}}{(1+n_p)} g_{{}_L}(\zeta_>,p)g_{{}_R} (\zeta_<,p) \\
&= \red{-}\frac{1}{\left\{K^{\text{in}}(p) - \left[K^{\text{in}}(p)\right]^\ast \right\}} \\
&\hspace{1cm}\times\frac{e^{\beta p^0 \zeta_0}}{(1+n_p)} \Big\{\Theta_{\text{SK}} (\zeta>\zeta_0) g_{{}_L} (\zeta,p) g_{{}_R} (\zeta_0,p) + \Theta_{\text{SK}} (\zeta<\zeta_0) g_{{}_L}(\zeta_0,p) g_{{}_R} (\zeta,p)\Big\} \ ,
\end{split}
\label{eq:BlkBlkgRgL}
\end{equation}
where $\Theta_{\text{SK}}$ denotes the theta function for which the ordering of the arguments is understood to be on the Schwinger-Keldysh contour. We can now use the definitions of $g_{\sL}(\zeta,p)$ and $g_{\sR}(\zeta,p)$ in Eq.\eqref{eq:DefgL} and Eq.\eqref{eq:DefgR} respectively to write
\begin{equation}
\begin{split}
\frac{e^{\beta p^0 \zeta_0}}{(1+n_p)} g_{{}_L}(\zeta_>,p)g_{{}_R} (\zeta_<,p) &= n_{p} e^{\beta p^0 \zeta_0}\ G^{\text{in}} (\zeta, p) G^{\text{in}} (\zeta_0,p)+(1+ n_{p}) e^{-\beta p^0 \zeta} \left[G^{\text{in}} (\zeta,p) G^{\text{in}} (\zeta_0,p)\right]^\ast\\
&\hspace{1cm} -e^{\beta p^0 (\zeta_0-\zeta)} \left[G^{\text{in}}(\zeta,p)\right]^\ast G^{\text{in}}(\zeta_0,p) \Big\{\Theta_{\text{SK}} (\zeta > \zeta_0) + n_{p}\Big\}\\
& \hspace{2.6cm} - G^{\text{in}}(\zeta,p) \left[G^{\text{in}}(\zeta_0,p)\right]^\ast \Big\{\Theta_{\text{SK}} (\zeta < \zeta_0) + n_{p}\Big\} \ .
\end{split}
\end{equation}
Substituting this result in Eq.(\eqref{eq:BlkBlkgRgL}), we get
\begin{equation}
\begin{split}
\mathbb{G}(\zeta|\zeta_0,p) &=  \red{-}\frac{1}{ \left\{ K^{\text{in}}(p) - \left[K^{\text{in}}(p)\right]^\ast \right\}}\\
&\qquad \times \bigg\{n_{p}\ e^{\beta p^0 \zeta_0}\ G^{\text{in}} (\zeta, p) G^{\text{in}} (\zeta_0,p)+(1+ n_{p})\ e^{-\beta p^0 \zeta} \left[G^{\text{in}} (\zeta,p) G^{\text{in}} (\zeta_0,p)\right]^\ast\\
&\hspace{4cm} -\big[\thetaSK (\zeta<\zeta_0)+n_p\big] \Gin(\zeta,p) \left[\Gin(\zeta_0,p)\right]^\ast\\
&\hspace{4cm} - \big[\thetaSK (\zeta>\zeta_0)+n_p\big] e^{\beta p^0(\zeta_0-\zeta)} \left[ \Gin(\zeta,p)\right]^\ast \Gin(\zeta_0,p) \bigg\} \ .
\end{split}
\label{eq:BlkBlkGinGinst}
\end{equation}

\subsection{Advanced and retarded bulk-to-bulk Green functions}\label{sec:Gret&Gadv}
In the previous sections, we calculated the bulk-to-bulk Green functions $\bbG(\zeta_2|\zeta_1,p)$ using the fact that it solved the equation of motion with a bulk delta source (see Eq.\eqref{eq:BlkBlkODE}). The boundary conditions we employed in doing so are normalisability conditions on both the left and the right boundaries of the grSK geometry under consideration. We shall now impose normalisability only on the left boundary, along with causality, and derive the corresponding bulk-to-bulk Green functions. We will find a retarded bulk-to-bulk Green function $\bbGR(\zeta_2|\zeta_1,p)$ and an advanced bulk-to-bulk Green function $\bbGA(\zeta_2|\zeta_1,p)$. Recall that causality (in real space) implies analyticity in the lower half of the complex $p^0$ plane for the advanced Green function $\bbGA(\zeta_2|\zeta_1,p)$ and in the upper half for the retarded Green function $\bbGR(\zeta_2|\zeta_1,p)$.
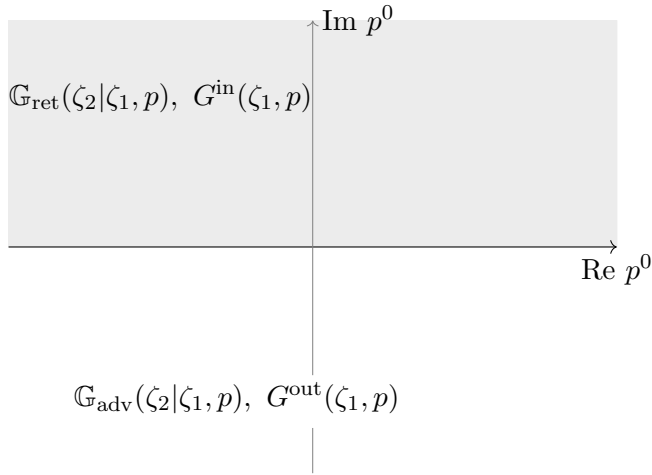
\begin{figure}
    \centering
    \begin{tikzpicture}[decoration={markings, 
        mark= at position 3.5cm with {\arrow{stealth}},
        mark= at position 8.0cm with {\arrow{stealth}},
        mark= at position 0.85 with {\arrow{stealth}}}
    ]
    \draw[lightgray!30,fill = lightgray!30] (-4,0)--(4,0)--(4,3)--(-4,3)--cycle;
    \draw[->] (-4,0) -- (4,0) coordinate (xaxis);
    \draw[help lines,->] (0,-1.7)--(0,3) coordinate (yaxis);
    \draw[help lines] (0,-3)--(0,-2.4) coordinate (yaxis);
    \node at (4,-0.3) {$\text{Re}\ p^0$};
    \node at (0.6,3) {$\text{Im}\ p^0$};
    \node at (-2, 2) {$\bbGR(\zeta_2|\zeta_1,p),\ \Gin(\zeta_1,p)$};
    \node at (-1, -2) {$\bbGA(\zeta_2|\zeta_1,p), \ G^{\rm out}(\zeta_1,p)$};
    \end{tikzpicture}
    \caption{Regions of analyticity of the various Green functions. The retarded bulk-to-bulk Green function $\bbGR(\zeta_2|\zeta_1,p)$ and the ingoing boundary-to-bulk Green function $\Gin(\zeta_1,p)$ are analytic in the shaded region (UHP). Meanwhile, the advanced bulk-to-bulk Green function $\bbGA(\zeta_2|\zeta_1,p)$ and the outgoing boundary-to-bulk Green function $G^{\rm out}(\zeta_1,p)$ are analytic in the unshaded region (LHP).}
    \label{fig:RegionAnalyticityGretGadv}
\end{figure}

Before we proceed to determine the explicit form of the advanced and the retarded Green functions, let us understand some of their general features. Both these Green functions satisfy the equation of motion with a bulk delta source (Eq.\eqref{eq:BlkBlkODE}). Thus when $\zeta_2 \neq \zeta_1$, both $\bbGR(\zeta_2|\zeta_1,p)$ and $\bbGA(\zeta_2|\zeta_1,p)$ must solve the homogeneous equation of motion, Eq.\eqref{eq:BlkBlkODE}. Therefore we conclude that both of them must be proportional to linear combinations of the homogeneous solutions when $\zeta_2 \neq \zeta_1$.  Since both $\bbGR(\zeta_2|\zeta_1,p)$ and $\bbGA(\zeta_2|\zeta_1,p)$ are normalisable on the left-boundary ($\zeta =0$), we have
\begin{equation}
    \bbG_{\text{ret, adv}}(\zeta_2|\zeta_1,p) \propto g_{\sR} (\zeta_<,p) \ ,
\end{equation}
where $g_{\sR}(\zeta,p)$ is the coefficient of the right-source $J_{R}$ in the Son-Teaney solution, defined in Eq.\eqref{eq:DefgR}. Here we are using the notation from Eq.\eqref{eq:DefZetaless} and Eq.\eqref{eq:DefZetagreat}.

\subsubsection{Retarded bulk-to-bulk Green function}
As discussed earlier, the retarded Green function is uniquely determined by the requirement that it is left-normalisable and analytic in the upper half of the complex $p^0$ plane.
We can write down a continuous function of the form
\begin{equation}
    \bbGR (\zeta_2|\zeta_1,p) = \frac{1}{\WR(\zeta_1,p)} \Gin(\zeta_>,p) g_{\sR} (\zeta_<,p) \ ,
    \label{eq:BlkBlkretAnsatz}
\end{equation}
where $\WR(\zeta_1,p)$ is a Wronskian factor that we will determine similarly as in Sec.(\ref{sec:WronskianEvaluation}). Note that this expresson is analytic in the upper half of the complex $p^0$ plane since $\Gin(\zeta_>,p)$ is analytic there and the poles of $\WR(\zeta_1,p)$ cancel that of $g_{\sR}(\zeta_<,p)$ (as we will see in a moment).

Integrating the equation of motion, Eq.\eqref{eq:BlkBlkODE} gives us a jump-discontinuity for the retarded Green function $\bbGR (\zeta_2|\zeta_1,p)$ of the form given in Eq.\eqref{eq:BlkBlkJumpCondition}. This jump condition determines the Wronskian factor $\WR(\zeta,p)$ to be of the form
\begin{equation}
    \begin{split}
        \WR(\zeta,p) &\equiv \red{-}\frac{2}{i \beta} r^{d-1} \left[g_{\sR} \left(\frac{\diff }{\diff \zeta} + \frac{\beta p^0}{2}\right) \Gin - \Gin \left(\frac{\diff }{\diff \zeta} + \frac{\beta p^0}{2}\right) g_{\sR} \right]\\
        &= \red{-}r^{d-1} \Big[g_{\sR} \ \mathbb{D}_+ \Gin - \Gin \ \mathbb{D}_+ g_{\sR} \Big] \ ,
    \end{split}
\end{equation}
where $\mathbb{D}_+$ is a differential operator that we defined earlier in Eq.\eqref{eq:DefDPlus}. We can now evaluate this Wronskian by using the constancy of $e^{\beta p ^0 \zeta} W(\zeta,p)$ that we proved earlier in Sec.(\ref{sec:WronskianConstancy}) for the Wronskian of $g_{\sR}(\zeta,p)$ and $g_{\sL}(\zeta,p)$. The reader can check that the same result holds for $\WR(\zeta,p)$ as well. The above expression also has the bare values of the conjugate fields of $\Gin$ and $g_{\sR}$. These get corrected due to counter-term contributions coming from holographic renormalisation. We will replace the conjugate fields appearing in the above expression by their renormalised versions. This doesn't change anything since the difference between the bare and the renormalised values of the conjugate fields are counter-terms linear in $\Gin$ and $g_{\sR}$. Therefore, using the boundary values of $\Gin$ and $g_{\sR}$ in Eq.\eqref{eq:LimBoundaryL_gRL} and Eq.\eqref{eq:LimBoundaryR_gRL} respectively along with those of the corresponding renormalised conjugate fields in Eq.\eqref{eq:GinConjRen} and Eq.\eqref{eq:LimBoundaryL_gRLRen} respectively, we get
\begin{equation}
    \begin{split}
        \lim_{\zeta \to 0} e^{\beta p^0 \zeta} \WR (\zeta,p) &= \red{-}\lim_{\zeta \to 0} \bigg\{ g_{\sR} \Big( r^{d-1} \mathbb{D}_+ \Gin \Big) \Big|_{\text{ren}} - \Gin \Big( r^{d-1} \mathbb{D}_+ g_{\sR} \Big)\Big|_{\text{ren}}\bigg\} = \red{-}\lim_{\zeta \to 0} g_{\sR}^{\pi,  \text{ren}}\\
        &= (1+n_p) \left\{\Kin (p) - \left[\Kin(p)\right]^\ast\right\} \ .
    \end{split}
\end{equation}
This gives the Wronskian $\WR(\zeta,p)$ to be
\begin{equation}
    \WR (\zeta,p) = \frac{(1+n_p) \left\{\Kin (p) - \left[\Kin(p)\right]^\ast\right\}}{e^{\beta p^0 \zeta}} \ .
\end{equation}

Substituting the above expression in Eq.\eqref{eq:BlkBlkretAnsatz}, we get the retarded bulk-to-bulk Green function $\bbGR (\zeta_2|\zeta_1,p)$ to be
\begin{equation}
    \bbGR (\zeta_2|\zeta_1,p) = \frac{e^{\beta p^0 \zeta_1}}{\Kin (p) - \left[\Kin(p)\right]^\ast} \Gin(\zeta_>,p)  \left\{G^{\text{in}}(\zeta_<,p) - e^{-\beta p^0 \zeta_<} \left[G^{\text{in}} (\zeta_<,p)\right]^\ast \right\}  \ ,
    \label{eq:BlkBlkretWronskianExplicit}
\end{equation}
where we have used the definition of $g_{\sR}(\zeta,p)$ given in Eq.\eqref{eq:DefgR}. Notice that the final factor in the above expression has poles in the upper half plane, but these are cancelled by the corresponding poles coming from the denominator. Thus $\bbGR (\zeta_2|\zeta_1,p)$ is indeed analytic in the upper half of the complex $p^0$ plane. Sometimes, we will find it  advantageous to write the above propagator as
\begin{equation}
    \begin{split}
        \bbGR (\zeta_2|\zeta_1,p) &= \red{-}\frac{1}{\Kin (p) - \left[\Kin(p)\right]^\ast}\\
        &\times \bigg\{\thetaSK (\zeta_2>\zeta_1) \Gin(\zeta_2,p)  \left[G^{\text{in}} (\zeta_1,p)\right]^\ast   \\
        &\quad+\thetaSK (\zeta_2<\zeta_1) e^{\beta p^0 (\zeta_1-\zeta_2)}   \left[G^{\text{in}} (\zeta_2,p)\right]^\ast \Gin(\zeta_1,p)   - e^{\beta p^0 \zeta_1}G^{\text{in}}(\zeta_1,p) G^{\text{in}}(\zeta_2,p)  \bigg\}  \ .
    \end{split}
    \label{eq:BlkBlkretExplicitThetafunc}
\end{equation}

\subsubsection{Advanced bulk-to-bulk Green function}
To evaluate the advanced bulk-to-bulk Green function $\bbGA (\zeta_2|\zeta_1,p)$, we will follow the same strategy we employed in the previous section. We can write down a continuous function of the form
\begin{equation}
    \bbGA (\zeta_2|\zeta_1,p) = \frac{1}{\WA(\zeta_1,p)} e^{-\beta p^0 \zeta_>} \left[\Gin(\zeta_>,p)\right]^\ast g_{\sR} (\zeta_<,p) \ ,
    \label{eq:BlkBlkadvAnsatz}
\end{equation}
where $\WA(\zeta,p)$ is a Wronskian factor to be determined by the jump condition in Eq.\eqref{eq:BlkBlkJumpCondition}. This jump condition determines the Wronskian factor $\WA(\zeta,p)$ to be of the form
\begin{equation}
    \begin{split}
        \WA(\zeta,p) &\equiv \red{-}\frac{2}{i \beta} r^{d-1} \left[g_{\sR} \left(\frac{\diff }{\diff \zeta} + \frac{\beta p^0}{2}\right) e^{-\beta p^0 \zeta}\left(\Gin\right)^\ast - e^{-\beta p^0 \zeta}\left(\Gin\right)^\ast \left(\frac{\diff }{\diff \zeta} + \frac{\beta p^0}{2}\right) g_{\sR} \right]\\
        &=  \red{-}g_{\sR} \Big[ r^{d-1} \mathbb{D}_+ e^{-\beta p^0 \zeta}\left(\Gin\right)^\ast \Big] \Big|_{\text{ren}} \red{+} e^{-\beta p^0 \zeta}\left(\Gin\right)^\ast \Big( r^{d-1} \mathbb{D}_+ g_{\sR} \Big)\Big|_{\text{ren}} \ ,
    \end{split}
\end{equation}
where in the second line we have renormalised the conjugate fields. We note again that this replacement does not change the Wronskian since the counter terms of the conjugate fields are proportional to the corresponding fields.
We can now use the constancy of $e^{\beta p^0 \zeta}\WA (\zeta,p)$ to evaluate the Wronskian. At the left-boundary, we get
\begin{equation}
    \begin{split}
        \lim_{\zeta \to 0} e^{\beta p^0 \zeta} \WA (\zeta,p) &= \red{-}\lim_{\zeta \to 0} \bigg\{ g_{\sR} \Big[ r^{d-1} \mathbb{D}_+ e^{-\beta p^0 \zeta}\left(\Gin\right)^\ast \Big] \Big|_{\text{ren}} - e^{-\beta p^0 \zeta}\left(\Gin\right)^\ast \Big( r^{d-1} \mathbb{D}_+ g_{\sR} \Big)\Big|_{\text{ren}}\bigg\}\\
        &= (1+n_p) \left\{\Kin (p) - \left[\Kin(p)\right]^\ast\right\} \ .
    \end{split}
\end{equation}
which gives 
\begin{equation}
    \WA (\zeta_1,p) = \frac{ (1+n_p) \left\{\Kin (p) - \left[\Kin(p)\right]^\ast\right\}}{e^{\beta p^0 \zeta_1}} \ .
\end{equation}
Substituting the above expression and the definition of $g_{\sR}(\zeta,p)$ from Eq.\eqref{eq:DefgR} in Eq.\eqref{eq:BlkBlkadvAnsatz}, we get
\begin{equation}
    \begin{split}
        \bbGA (\zeta_2|\zeta_1,p) &= \frac{e^{\beta p^0 \zeta_1}}{\Kin (p) - \left[\Kin(p)\right]^\ast} e^{-\beta p^0 \zeta_>} \left[\Gin(\zeta_>,p)\right]^\ast\\
        & \hspace{2cm} \times  \left\{G^{\text{in}}(\zeta_<,p) - e^{-\beta p^0 \zeta_<} \left[G^{\text{in}} (\zeta_<,p)\right]^\ast \right\}  \ .
    \end{split}
    \label{eq:BlkBlkadvWronskianExplicit}
\end{equation}
We have already seen in the previous section that the poles of $g_{\sR} (\zeta_>,p)$ cancel the corresponding poles from the Wronskian in the denominator. Notice also that $e^{-\beta p^0 \zeta_>} \left[\Gin(\zeta_>,p)\right]^\ast$ is analytic in the lower half of the complex $p^0$ plane. Thus $\bbGA (\zeta_2|\zeta_1,p)$ is indeed analytic in the lower half of the complex $p^0$ plane and deserves to be called the advanced bulk-to-bulk Green function. More explicitly, in terms of the theta functions on the SK contour, the advanced Green function takes the form
\begin{equation}
    \begin{split}
        \bbGA (\zeta_2|\zeta_1,p) &=\frac{1}{\Kin (p) - \left[\Kin(p)\right]^\ast}\\
        &\times \bigg\{ \thetaSK(\zeta_2>\zeta_1)e^{\beta p^0 (\zeta_1-\zeta_2)} \left[\Gin(\zeta_2,p)\right]^\ast  G^{\text{in}}(\zeta_1,p)  \\
        &\quad+ \thetaSK(\zeta_2<\zeta_1)  G^{\text{in}}(\zeta_2,p)  \left[\Gin(\zeta_1,p)\right]^\ast   -  e^{-\beta p^0 \zeta_2}\left[G^{\text{in}} (\zeta_2,p)  G^{\text{in}} (\zeta_1,p)\right]^\ast \bigg\}  \ .
    \end{split}
    \label{eq:BlkBlkadvExplicitThetafunc}
\end{equation}
Notice that the retarded and the advanced bulk-to-bulk Green functions transform into each other under field point-source point reversal and momentum reversal. This can be seen by performing this transformation on Eq.\eqref{eq:BlkBlkWronskianExplicit} as follows
\begin{equation}
    \bbGA (\zeta_1|\zeta_2,-p) = \red{-}\frac{e^{-\beta p^0 \zeta_2}}{\Kin (p) - \left[\Kin(p)\right]^\ast} e^{\beta p^0 \zeta_>} \Gin(\zeta_>,p) \left\{\left[G^{\text{in}}(\zeta_<,p) \right]^\ast - e^{\beta p^0 \zeta_<} G^{\text{in}} (\zeta_<,p) \right\}  \ ,
\end{equation}
where we have used the results $\Gin(\zeta,-p) = \left[\Gin(\zeta,p) \right]^\ast$ and $\Kin(-p) = \left[\Kin(p)\right]^\ast$. The above expression can be simplified as
\begin{equation}
    \bbGA (\zeta_1|\zeta_2,-p) = \red{-}\frac{e^{\beta p^0 \zeta_1}}{\Kin (p) - \left[\Kin(p)\right]^\ast}  \Gin(\zeta_>,p) \left\{e^{-\beta p^0 \zeta_<}\left[G^{\text{in}}(\zeta_<,p) \right]^\ast -  G^{\text{in}} (\zeta_<,p) \right\}  \ ,
\end{equation}
where we have used $\zeta_<+\zeta_> = \zeta_1+\zeta_2$, which follows from the definitions of $\zeta_<$ and $\zeta_>$ in Eq.\eqref{eq:DefZetaless} and Eq.\eqref{eq:DefZetagreat} respectively. Thus we find, as claimed earlier, that
\begin{equation}
    \bbGA(\zeta_1|\zeta_2,-p) = \bbGR(\zeta_2|\zeta_1,p) \ .
    \label{eq:BlkBlkretadvReciprocity}
\end{equation}

\subsubsection{Manifestly reciprocal representation of $\bbG(\zeta_2|\zeta_1,p)$}\label{sec:BlkBlkReciprocalRepresentation}
We end this section by noting that the bulk-to-bulk Green function $\bbG(\zeta_2|\zeta_1,p)$ that is normalisable on both the left and the right boundaries can be written as a linear combination of the advanced and the retarded bulk-to-bulk Green functions as follows
\begin{equation}
    \bbG(\zeta_2|\zeta_1,p) = -n_p\bbGR(\zeta_2|\zeta_1,p) + (1+n_p) \bbGA(\zeta_2|\zeta_1,p) \ .
\end{equation}
Notice that this is a manifestly reciprocal (Eq.\ref{eq:BlkBlkreciprocity}) representation of the bulk-to-bulk Green function $\bbG(\zeta_2|\zeta_1,p)$, since under reciprocity $n_p \mapsto -(1+n_p)$ and $\bbGA(\zeta_1|\zeta_2,-p) \mapsto \bbGR(\zeta_2|\zeta_1,p)$ as we showed in Eq.\eqref{eq:BlkBlkretadvReciprocity}.

\subsection{Corrections to the CFT expectation values}\label{sec:CFTexpecValuesCorrections}

As noted in the main text, the first-order correction $\phi_{(1)}$ to the bulk solution is
\begin{equation}
    \phi_{(1)} (\zeta,p) = \oint_{\zeta_0} \ \mathbb{G}(\zeta|\zeta_0,p) \ \mathcal{J}^{\mathfrak{B}}_{(1)} (\zeta_0,p) \ ,
\end{equation}
where $\mathcal{J}^{\mathfrak{B}}_{(1)}(\zeta,p)$ is the bulk source at first order defined in Eq.\eqref{eq:Phi1BulkSource}. 
Given this correction, we would now like to compute the correction to the expectation values of the dual CFT operator $\mathcal{O}$. These expectation values are given by boundary limits of the renormalised conjugate field (see Eq.~\eqref{eq:ExpecvalfromRenConjField}), which can be expanded as
\begin{equation}
    \pi^{\text{ren}} = \pi^{\text{ren}}_{(0)}+ \lambda_{3\text{B}} \pi^{\text{ren}}_{(1)}+ \lambda^2_{3 \rm B} \pi^{\text{ren}}_{(2)} + \ldots \ .
\end{equation}
We have already computed the leading-order result $\pi^{\text{ren}}_{(0)}$ in Eq.\eqref{eq:ConjugateFieldLeadingRen} whose boundary limits yielded the expectation values given in Eq.\eqref{eq:CFTExpValLeadingRL}. The sub-leading correction is given by
\begin{equation}
    \begin{split}
        \pi_{(1)}^{\text{ren}}(\zeta,p) &\equiv - r^{d-1} \mathbb{D}_+ \phi_{(1)}\Big|_{\text{ren}}\ = \oint_{\zeta_0} \left[- r^{d-1} \mathbb{D}_+ \ \mathbb{G}(\zeta|\zeta_0,p)\right]\bigg|_{\text{ren}} \ \mathcal{J}^{\mathfrak{B}} (\zeta_0,p)\ .
    \end{split}
    \label{eq:ConjugateFieldSubleadRen}
\end{equation}

We will begin by evaluating the renormalised conjugate field associated with the bulk-to-bulk Green function at the left and the right boundaries. We obtain 
\begin{equation}
    \begin{split}
        -\lim_{\zeta \to 0} r^{d-1} \mathbb{D}_+ \mathbb{G}(\zeta|\zeta_0,p)|_{\text{ren}} &=  \red{-}\frac{e^{\beta p^0 \zeta_0}}{(1+ n_p) \left[ K^{\text{in}} - (K^{\text{in}})^\ast\right]} g_{\sL} (\zeta_0,p)\lim_{\zeta \to 0} g^{\pi,{\text{ren}}}_{\sR}(\zeta,p)\\
        &= e^{\beta p^0 \zeta_0} g_{\sL} (\zeta_0,p) = - g_{\sL} (\zeta_0,-p) \ , 
    \end{split}
\end{equation}
and
\begin{equation}
    \begin{split}
        -\lim_{\zeta \to 1} r^{d-1} \mathbb{D}_+ \mathbb{G}(\zeta|\zeta_0,p)|_{\text{ren}} &=  \red{-} \frac{e^{\beta p^0 \zeta_0}}{(1+ n_p) \left[ K^{\text{in}} - (K^{\text{in}})^\ast\right]} g_R (\zeta_0,p)\lim_{\zeta \to 1} g^{\pi,{\text{ren}}}_{\sL}(\zeta,p)\\
        &= e^{-\beta p^0 (1-\zeta_0)} g_{\sR} (\zeta_0,p) =-g_{\sR}(\zeta_0,-p)\ .
    \end{split}
\end{equation}
Here, we have used the limits from Eq.\eqref{eq:LimBoundaryL_gRLRen} and Eq.\eqref{eq:LimBoundaryR_gRLRen} along with the momentum-reversal properties of $g_{\sR}(\zeta,p)$ and $g_{\sL}(\zeta,p)$ given in Eq.\eqref{eq:MomRevgRL}. Note that these are the analogues of the LSZ reduction for the exterior EFT in that the boundary-to-bulk propagators can be obtained from the bulk-to-bulk propagators by applying appropriate derivatives.

Now, the SK expectation values of the CFT operator $\mathcal{O}$ can also be expanded as
\begin{equation}
    \begin{split}
        \langle \mathcal{O}_{\sR} \rangle &= \langle \mathcal{O}_{\sR} \rangle_{(0)}+  \langle \mathcal{O}_{\sR} \rangle_{(1)} +  \langle \mathcal{O}_{\sR} \rangle_{(2)}+ \ldots \ ,\\
        \langle \mathcal{O}_{\sL} \rangle &= \langle \mathcal{O}_{\sL} \rangle_{(0)}+  \langle \mathcal{O}_{\sL} \rangle_{(1)} +  \langle \mathcal{O}_{\sL} \rangle_{(2)}+ \ldots \ .
    \end{split}
\end{equation}
The leading-order answers $\langle \mathcal{O}_{\sR, \sL} \rangle_{(0)}$ were computed in Eq.\eqref{eq:CFTExpValLeadingRL}. At the next-to-leading order, we obtain
\begin{equation}
    \begin{split}
        \langle \mathcal{O}_{\sR} \rangle_{(1)} &= \lambda_{3\text{B}}\oint_{\zeta_0}  g_{\sR} (\zeta_0,-p) \mathcal{J}^{\mathfrak{B}}(\zeta_0,p)\ ,\\
        \langle \mathcal{O}_{\sL} \rangle_{(1)} &= \lambda_{3\text{B}}\oint_{\zeta_0}  g_{\sL} (\zeta_0,-p) \mathcal{J}^{\mathfrak{B}}(\zeta_0,p)\ .
    \end{split}
    \label{eq:CFTexpecNLO_RL}
\end{equation}
This is a simple and intuitive result: the effect of the bulk source shows up in the boundary expectation values through a momentum-reversed boundary-to-bulk Green function. In other words, the kernel of the integral operator that produces the boundary CFT one-point functions given the bulk sources is proportional to the momentum-reversed boundary-to-bulk Green functions.

We will now find explicit expressions for the above one-point functions and also check that the Schwinger-Keldysh collapse and KMS conditions hold. To this end, we begin with the leading-order solution $\phi_{(0)}$ written in the past-future basis:
\begin{equation}
    \phi_{(0)}(\zeta, k) =- G^{\text{in}} (\zeta, k) J_{\Fb}(k)+ e^{\beta k^0 } G^{\text{out}}(\zeta, k)  J_{\Pb}(k) \ ,
\end{equation}
in terms of which the bulk source $\mathcal{J}^{\mathfrak{B}}$ takes the form
\begin{equation}
    \begin{split}
        \mathcal{J}^{\mathfrak{B}}(\zeta_0,p) &=  \red{-} \int_{k_1} \int_{k_2} (2\pi)^d \delta^{(d)}(k_1 + k_2 -p) \Bigg\{\frac{1}{2}J_{\Fb} (k_1) J_{\Fb}(k_2) G^{\text{in}} (\zeta_0, k_1) G^{\text{in}} (\zeta_0, k_2)\\
        &\hspace{4cm} -J_{\Fb} (k_1) J_{\Pb}(k_2) G^{\text{in}} (\zeta_0, k_1) e^{\beta k_2^0 (1-\zeta_0)} \left[G^{\text{in}}(\zeta_0, k_2)\right]^\ast\\
        &\hspace{4cm}  + \frac{1}{2}J_{\Pb} (k_1) J_{\Pb}(k_2) e^{\beta p^0 (1-\zeta_0)} \left[G^{\text{in}}(\zeta_0, k_1)G^{\text{in}}(\zeta_0, k_2)\right]^\ast\Bigg\} \ .
    \end{split}
    \label{BulkSourcePF}
\end{equation}
Substituting the above expression in Eq.\eqref{eq:CFTexpecNLO_RL} and using the expressions for $g_{\sR, \sL} (\zeta,p)$ from Eq.\eqref{eq:DefgL} and Eq.\eqref{eq:DefgR}, we get
\begin{equation}
    \begin{split}
        \big\langle \mathcal{O}_{\sR} -\mathcal{O}_{\sL} \big\rangle_{(1)} &= -\lambda_{3\text{B}} \int_{k_1} \int_{k_2} (2\pi)^d \delta^{(d)}(k_1 + k_2 -p)  \oint_{\zeta_0} \\
        &\times \Bigg\{-J_{\Fb} (k_1) J_{\Pb}(k_2) e^{\beta k_2^0 (1-\zeta_0)}     \ \left[G^{\text{in}}(\zeta_0,p)\right]^\ast  G^{\text{in}} (\zeta_0, k_1)  \left[ G^{\text{in}}(\zeta_0, k_2)\right]^\ast\\
        &\qquad + \frac{1}{2}J_{\Pb} (k_1) J_{\Pb}(k_2) e^{\beta p^0(1- \zeta_0)}\ \left[G^{\text{in}}(\zeta_0,p) G^{\text{in}}(\zeta_0, k_1)G^{\text{in}}(\zeta_0, k_2)\right]^\ast\Bigg\} \ .
    \end{split}
\end{equation}
\begin{equation}
    \begin{split}
        \big\langle \mathcal{O}_{\sR}  - e^{-\beta p^0}\mathcal{O}_{\sL} \big\rangle_{(1)} &= -\lambda_{3\text{B}} \int_{k_1} \int_{k_2} (2\pi)^d \delta^{(d)}(k_1 + k_2 -p)  \oint_{\zeta_0} \\
        &\times  \Bigg\{\frac{1}{2}J_{\Fb} (k_1)J_{\Fb}(k_2) e^{-\beta p^0 (1-\zeta_0)} G^{\text{in}} (\zeta_0,p) G^{\text{in}} (\zeta_0, k_1) G^{\text{in}} (\zeta_0, k_2)\\
        &\qquad -J_{\Fb} (k_1) J_{\Pb}(k_2)   e^{\beta k_1^0 (\zeta_0-1)} \ G^{\text{in}} (\zeta_0,p)\ G^{\text{in}} (\zeta_0, k_1)  \left[ G^{\text{in}}(\zeta_0, k_2)\right]^\ast \Bigg\} \ .\\
    \end{split}
\end{equation}
We have only given the above combinations of the expectation values since we aim to check the Schwinger-Keldysh collapse and KMS conditions. From the above expressions it is clear that 
\begin{equation}
    \big\langle \mathcal{O}_{\sR}- \mathcal{O}_{\sL} \big\rangle_{(1)}\Big|_{J_{\Pb} =0} = 0 \ , \quad \text{and} \quad
    \big\langle \mathcal{O}_{\sR}-e^{-\beta p^0} \mathcal{O}_{\sL} \big\rangle_{(1)} \Big|_{J_{\Fb} =0} = 0 \ .
\end{equation}
The first of the above equalities shows that the CFT expectation values obey the Schwinger-Keldysh collapse conditions. The second equality verifies the Kubo-Martin-Schwinger relations for the boundary CFT at inverse temperature $\beta$.

\subsection{Vertices at finite temperature}

The Feynman rules we have presented in table~\ref{tab:Vertices} for the three-point and the four-point vertices are exactly those of a unitary theory written in the past-future basis. To see this, consider, for example, the simple example of a unitary theory with a three-point coupling. The interaction term in the action takes the form
\begin{equation}
    S_{3} = -\frac{\lambda_{3 \rm B}}{3!} \int_{k_1} \int_{k_2} \int_{k_3}  (2\pi)^d \delta^{(d)} \left(\sum_{i=1}^{3} k_i\right) \left(\prod_{i=1}^3 J_{\sR} (k_i) - \prod_{i=1}^3 J_{\sL} (k_i) \right)\ .
\end{equation}
This term when written in the past-future basis takes the form
\begin{equation}
    S_{n} = \frac{\lambda_{3 \rm B}}{2!} \int_{k_1} \int_{k_2} \int_{k_3}(2\pi)^d \delta^{(d)}\left(\sum_{i=1}^{3} k_i\right) \left[J_{\Fb}(k_1) J_{\Pb}(k_2) J_{\Pb}(k_3) +\frac{n_{-k_2} n_{-k_3}}{n_{k_1}} J_{\Pb}(k_1) J_{\Fb}(k_2) J_{\Fb}(k_3)\right] \ .
\end{equation}
Note that this exactly reproduces the vertices given in table~\ref{tab:Vertices}.

\subsection{Comparison with the textbook by Gelis}

We will now compare our Feynman rules with those presented in \cite{Gelis:1997zv} for thermal field theory. To this end, we will introduce a general basis $\{J_{p}, J_{f}\}$ defined as
\begin{equation}
    \begin{pmatrix}
        J_{p}(k)\\
        J_{f}(k)
    \end{pmatrix}
    =
    \begin{pmatrix}
        a(k)&-a(k)\\
        b(k)\frac{-n_{-k}}{a(-k)}&b(k)\frac{-n_k}{a(-k)}
    \end{pmatrix}
    \begin{pmatrix}
        J_{\sR}(k)\\
        J_{\sL}(k)
    \end{pmatrix}
     \ ,
\end{equation}
which will reduce to our past-future basis for certain choices of the functions $a$ and $b$ and likewise for the basis in \cite{Gelis:1997zv}.

Here $a(k)$ and $b(k)$ are arbitrary functions of the zeroth component of the momentum. Furthermore, we work in a basis, where the propagators are given by
\begin{equation}
    \begin{tikzpicture}[scale=1.4]
        \draw[dashed] (-0.5,0)--(0.5,0);  
        \diode{0}{0}{0}{-1.7};
        \node at (0,-1.7) {$\bullet$};
        \node at (0.2,-1.7) {$\zeta$};
        \node at (0,0) {$\times$};
        \node at (-0.3,-0.5) {$k$};
        \node at (2.65, -0.75) {$  = c(k)\Gin(\zeta,k) J_{f}(k) \ ,$};
    \end{tikzpicture}
\hspace{2cm}
    \begin{tikzpicture}[scale =1.4]
        \draw[dashed] (-0.5,0)--(0.5,0);  
        \diode{0}{-1.7}{0}{0};
        \node at (0,-1.7) {$\bullet$};
        \node at (0.2,-1.7) {$\zeta$};
        \node at (0,0) {$\times$};
        \node at (-0.3,-0.5) {$k$};
        \node at (2.5, -0.75) {$ = d(k) G^{\rm out}(\zeta,k ) J_{p}(k) \ ,$};
    \end{tikzpicture}
\end{equation}
\begin{equation}
    \begin{tikzpicture}[scale=1.4]
        \diode{-1}{0}{1}{0};
        \node at (-0.3,-0.2) {$k$};
        \node at (-1,-0.2) {$\zeta_1$};
        \node at (1,-0.2) {$\zeta_2$};
        \node at (-1,0) {$\bullet$};
        \node at (1,0) {$\bullet$};
        \node at (3.5,0) {$ = i b(k)c(k)d(-k)\bbGR (\zeta_2|\zeta_1,k) \ .$};
    \end{tikzpicture}
\end{equation}
Once again, $c(k)$ and $d(k)$ are arbitrary functions of the zeroth component of the momentum. 

\begin{table}[H]
    \centering
    \begin{tabular}{|c|c|}
 \hline     & \\
       \begin{tikzpicture}[scale=1,baseline=(current bounding box.center)]]
        \diodearrow{0}{0}{0}{1}
        \node at (0,0) {$\bullet$};
        \node at (0.4,0.8) {$k_1$};
        \diodearrow{0}{0}{-1}{-0.5}
        \node at (-1,0) {$k_2$};
        \semicap{1}{-0.5}{0}{0}
        \node at (1,0) {$k_3$};
    \end{tikzpicture}  &  $i \lambda_{3\rm B} \frac{n_{k_1}n_{k_2}}{a(k_1)d(k_1)a(k_2)d(k_2)} \frac{a(-k_3)}{n_{-k_3}b(k_3)c(k_3)} $  \\   & \\
    \hline   & \\
    \begin{tikzpicture}[scale=1,baseline=(current bounding box.center)]]
        \diodearrow{0}{0}{0}{1}
        \node at (0,0) {$\bullet$};
        \node at (0.4,0.8) {$k_1$};
        \semicap{-1}{-0.5}{0}{0}
        \node at (-1,0) {$k_2$};
        \semicap{1}{-0.5}{0}{0}
        \node at (1,0) {$k_3$};
        \end{tikzpicture} & $ -i \lambda_{3\rm B}  \frac{1}{a(k_1)d(k_1)} \frac{a(-k_2)}{b(k_2)c(k_2)} \frac{a(-k_3)}{b(k_3)c(k_3)} $\\   & \\
    \hline   & \\
    \begin{tikzpicture}[scale=0.8,baseline=(current bounding box.center)]]
        \diodearrow{0}{0}{1}{1}
        \node at (0,0) {$\bullet$};
        \node at (0.9,0.4) {$k_1$};
        \diodearrow{0}{0}{-1}{1}
        \node at (-0.3,0.8) {$k_2$};
        \diodearrow{0}{0}{-1}{-1}
        \node at (-0.8,-0.35) {$k_3$};
        \semicap{1}{-1}{0}{0}
        \node at (0.5,-0.8) {$k_4$};
        \end{tikzpicture} & $  -i \lambda_{4\rm B}  \frac{n_{k_1}n_{k_2}n_{k_3}}{a(k_1)d(k1)a(k_2)d(k_2)a(k_3)d(k_3)} \frac{a(-k_4)}{n_{-k_4}b(k_4)c(k_4)} $\\   & \\
    \hline   & \\
    \begin{tikzpicture}[scale=0.8,baseline=(current bounding box.center)]]
        \diodearrow{0}{0}{1}{1}
        \node at (0,0) {$\bullet$};
        \node at (0.9,0.4) {$k_1$};
        \diodearrow{0}{0}{-1}{1}
        \node at (-0.3,0.8) {$k_2$};
        \semicap{-1}{-1}{0}{0}
        \node at (-0.8,-0.35) {$k_3$};
        \semicap{1}{-1}{0}{0}
        \node at (0.5,-0.8) {$k_4$};
\end{tikzpicture}    & $i \lambda_{4\rm B} \frac{1}{n_{-k_3-k_4}} \frac{n_{k_1}n_{k_2}}{a(k_1)d(k_1)a(k_2)d(k_2)} \frac{a(-k_3)}{b(k_3)c(k_3)} \frac{a(-k_4)}{b(k_4)c(k_4)}$\\  & \\
    \hline   & \\
    \begin{tikzpicture}[scale=0.8,baseline=(current bounding box.center)]]
        \diodearrow{0}{0}{1}{1}
        \node at (0,0) {$\bullet$};
        \node at (0.9,0.4) {$k_1$};
        \semicap{-1}{1}{0}{0}
        \node at (-0.3,0.8) {$k_2$};
        \semicap{-1}{-1}{0}{0}
        \node at (-0.8,-0.35) {$k_3$};
        \semicap{1}{-1}{0}{0}
        \node at (0.5,-0.8) {$k_4$};
        \end{tikzpicture} & $ -i \lambda_{4\rm B} \frac{1}{a(k_1)d(k_1)} \frac{a(-k_2)}{b(k_2)c(k_2)} \frac{a(-k_3)}{b(k_3)c(k_3)} \frac{a(-k_4)}{b(k_4)c(k_4)} $\\   & \\
    \hline
    \end{tabular}
    \caption{Feynman rules for the three-point and the four-point vertices.}
    \label{tab:VerticesGeneralabc}
\end{table}

With these normalisations, the three-and four-point vertices are given in table \ref{tab:VerticesGeneralabc}.

More generally, for an $n$-point function, we have the following rule for the vertex:
\begin{equation}
    i \lambda_{n \rm B} \frac{1}{n_{k_{\triangleright}}} \prod_{i \in |} \frac{a(-k_i)}{b(k_i)c(k_i)} \prod_{j \in \triangleright} \frac{-n_{k_j}}{a(k_j)d(k_j)} \ ,
\end{equation}
which can also be written as
\begin{equation}
    -i \lambda_{n \rm B} \frac{1}{n_{-k_{\triangleright}}} \prod_{i \in |} \frac{a(-k_i)}{b(k_i)c(k_i)} \prod_{j \in \triangleright} \frac{n_{-k_j}}{a(k_j)d(k_j)} \ .
\end{equation}

To match the above expression with our result in the main text, put $a(k)=-n_k$, $b(k)=-(1+n_{k})$, $c(k)=\frac{1}{1+n_k}$, $d(k)=1$. To match with the textbook by Gelis, put  put $a(k)=-n_k$, $b(k)=c(k)=d(k)=1$. Note that the textbook quotes rules for computing correlators, while we write rules for computing generating functionals. Thus before matching, one has to do a momentum reversal as well, i.e., $k \mapsto -k$.

\addcontentsline{toc}{section}{References}
\bibliographystyle{JHEP}

\bibliography{OpenEFTHawkRad}

\end{document}